\documentclass{emulateapj}

\def\hmpc{\rm \,h^{-1}\,Mpc}

\slugcomment{Accepted for publication on ApJS, COSMOS Special Issue}

\shorttitle{Galaxy morphology and 
large-scale structure at $z\sim 0.7$}
\shortauthors{Guzzo, et al.}

\begin{document}


\title{The Cosmic Evolution Survey (COSMOS): 
a large-scale structure at $z=0.73$ and the relation of
galaxy morphologies to local environment
} 


\author{L. Guzzo\altaffilmark{1,*}, P. Cassata\altaffilmark{2,3}, 
%
A. Finoguenov\altaffilmark{4},
R. Massey\altaffilmark{5},
N.Z. Scoville\altaffilmark{5,16},
P. Capak\altaffilmark{5},
R.S. Ellis\altaffilmark{5},
B. Mobasher\altaffilmark{6},
Y. Taniguchi\altaffilmark{7,8},
D. Thompson\altaffilmark{9,5},
%
M. Ajiki\altaffilmark{7},
H. Aussel\altaffilmark{10,16},
H. B\"ohringer\altaffilmark{4},
M. Brusa\altaffilmark{4},
D. Calzetti\altaffilmark{6},
A. Comastri\altaffilmark{11},
A. Franceschini\altaffilmark{3},
G. Hasinger\altaffilmark{4},
M.M. Kasliwal\altaffilmark{5},
M.G. Kitzbichler\altaffilmark{12},
J.-P. Kneib\altaffilmark{13}, 
A. Koekemoer\altaffilmark{6},
A. Leauthaud\altaffilmark{13}, 
H. J. McCracken\altaffilmark{14},
T. Murayama\altaffilmark{7},
T. Nagao\altaffilmark{7},
J. Rhodes\altaffilmark{15,5}, 
D.B. Sanders\altaffilmark{16},
S. Sasaki\altaffilmark{7},
Y. Shioya\altaffilmark{7},
L. Tasca\altaffilmark{12},
J.E. Taylor\altaffilmark{5}
}
\altaffiltext{1}{INAF - Osservatorio Astronomico di Brera, Via Bianchi 46,
I-23807, Merate (LC), Italy}
\altaffiltext{2}{INAF - Istituto di Astrofisica Spaziale e Fisica Cosmica, 
Sezione di Milano, Via Bassini 15, 20133, Milano, Italy}
\altaffiltext{3}{Dipartimento di Astronomia, Universit\`a di Padova,
Vicolo dell'Osservatorio 3, I-35100 Padova, Italy}
\altaffiltext{4}{Max-Planck Institut f\"ur Extraterrestrische Physik,
Giessenbach Str., Garching, Germany} 
\altaffiltext{5}{California Institute of Technology, MC 105-24, 1200 East
California Boulevard, Pasadena, CA 91125}
\altaffiltext{6}{Space Telescope Science Institute, 3700 SanMartin
Drive, Baltimore, MD 21218}
\altaffiltext{7}{Astronomical Institute, Graduate School of Science,
         Tohoku University, Aramaki, Aoba, Sendai 980-8578, Japan}
\altaffiltext{8}{Physics Department, Graduate School of Science, Ehime
University, 2-5 Bunkyou, Matuyama, 790-8577, Japan} 
\altaffiltext{9}{Large  Binocular Telescope Observatory, 933 N. Cherry
Ave.,  Tucson, AZ 85721-0065} 
\altaffiltext{10}{Service d'Astrophysique, CEA/Saclay, 91191
  Gif-sur-Yvette, France}
\altaffiltext{11}{INAF - Osservatorio Astronomico di Bologna, Via
  Ranzani, Bologna, Italy}
\altaffiltext{12}{Max-Planck-Institut f\"ur Astrophysik, D-85748
  Garching bei M\"unchen, Germany}
\altaffiltext{13}{Laboratoire d'Astrophysique de Marseille, BP 8, Traverse
du Siphon, 13376 Marseille Cedex 12, France}

\altaffiltext{14}{Institut d'Astrophysique de Paris, UMR7095 CNRS,
Universit\`e Pierre et Marie Curie, 98 bis Boulevard Arago, 75014
Paris, France} 
\altaffiltext{15}{Jet Propulsion Lab, 4800 Oak Grove Drive, Pasadena,
  CA 91109}
\altaffiltext{16}{University of Hawaii, 2680 Woodlawn Dr., Honolulu,
  HI 96822}
\altaffiltext{*}{Visiting scientist, ESO and MPA/MPE, Garching, Germany}
\altaffiltext{$\dagger$}{Based on observations with the NASA/ESA {\em Hubble
Space Telescope}, obtained at the Space Telescope Science Institute,
which is operated by AURA Inc, under NASA contract NAS 5-26555; also
based on data collected using: the Subaru Telescope, which is operated
by the National Astronomical Observatory of Japan; the XMM-Newton, an
ESA science mission with instruments and contributions directly funded
by ESA Member States and NASA; the European Southern Observatory,
Chile; Kitt Peak National Observatory, Cerro Tololo Inter-American
Observatory, and the National Optical Astronomy Observatory, which are
operated by the Association of Universities for Research in Astronomy,
Inc. (AURA) under cooperative agreement with the National Science
Foundation; the National Radio Astronomy Observatory which is a
facility of the National Science Foundation operated under cooperative
agreement by Associated Universities, Inc ; and MegaPrime/MegaCam, a
joint project of CFHT and CEA/DAPNIA, at the the Canada-France-Hawaii
Telescope operated by the National Research Council of Canada, the
Centre National de la Recherche Scientifique de France and the
University of Hawaii. Based in part on data products produced at TERAPIX
and CADC.}




\begin{abstract}
  We have identified a large-scale structure at $z\simeq 0.73$ in the
  COSMOS field, coherently described by the distribution of galaxy
  photometric redshifts, an ACS weak-lensing convergence map and the
  distribution of extended X-ray sources in a mosaic of XMM
  observations.  The main peak seen in these maps corresponds to a
  rich cluster with $T_X= 3.51_{-0.46}^{ +0.60}$ keV and $L_X=1.56\pm
  0.04 \times 10^{44}$ erg s$^{-1}$ ($[0.1-2.4]$ keV band).  We
  estimate an X-ray mass within $r_{500}$ corresponding to
  $M_{500}\simeq 1.6 \times10^{14} $ M$_\sun$ and a total lensing mass
  (extrapolated by fitting a NFW profile) $M_{NFW}=
  (6\pm3)\times10^{15}M_\sun$.  We use an automated morphological
  classification of all galaxies brighter than $I_{AB}=24$ over the
  structure area to measure the fraction of early-type objects as a
  function of local projected density $\Sigma_{10}$, based on
  photometric redshifts derived from ground-based deep multi-band
  photometry.  We recover a robust morphology-density relation at this
  redshift, indicating, for comparable local densities, a smaller
  fraction of early-type galaxies than today.  Interestingly, this
  difference is less strong at the highest densities and becomes more
  severe in intermediate environments.  We also find, however, local
  ``inversions'' of the observed global relation, possibly driven by
  the large-scale environment.  In particular, we find direct
  correspondence of a large concentration of disk galaxies to (the
  colder side of) a possible shock region detected in the X-ray
  temperature map and surface brightness distribution of the dominant
  cluster. We interpret this as potential evidence of shock-induced
  star formation in existing galaxy disks, during the ongoing merger
  between two sub-clusters.  Our analysis reveals the value of
  combining various measures of the projected mass density to locate
  distant structures and their potential for understanding the
  physical processes at work in the transformation of galaxy
  morphologies.
\end{abstract}

\section{Introduction}
The gravitational instability paradigm assumes that large-scale
structure in the Universe evolves in a hierarchical fashion, forming
larger and larger systems via the assembly of smaller sub-units.  It
is natural to think that  the development of this structure should have a
significant effect on the properties of galaxies that we observe today and
that a strong correlation between these and the surrounding
environment should be in general observed.
One of the most evident correlations of galaxy
properties with the environment in the present-day Universe is the
morphology-density (MD) relation: early-type galaxies, i.e.~$E$ and $S0$
({\it spheroidal}) gas-poor objects are preferentially found in dense environments, such as
groups and clusters
(Oemler 1974, Dressler 1980, Giovanelli, Haynes \& Chincarini 1986).
Measurements now exist also at redshifts up to unity, that show a
similar relationship to be already in place at earlier times
(Dressler et al. 1997, Andreon 1998, Smith et al. 2005, Postman et al. 2005).

Two competing 
scenarios can explain the MD relation.
On one side, it is tempting to associate its origin 
to the build-up of the hierarchical dark-matter
skeleton of the Universe.  If mergers played a role in the early construction
of the Hubble sequence, with (some or all) early-type galaxies
plausibly produced from the merging of gas-rich sub-systems, we would
expect
the cross-section for this process to depend strongly on the typical
velocity dispersion of the hosting structure, which in turn depends on
the mass scale of the environment where the galaxy lives.
Equally important effects on the evolution of the galaxies' stellar
and gaseous components may also be provided by other environmental
effects in rich clusters, like ram-pressure stripping by the dense
intra-cluster medium (ICM) (e.g. Gunn \& Gott 1972,
Quilis et al. 2000), close-encounters (Barnes 1992) and ``harassment''
via multiple weak encounters (Moore et al. 1996), or ``strangulation''
(Larson et al. 1980).   
This seemingly natural
picture would point to a {\it nurture} origin for the MD relation:
that it is the product of one or more evolutionary processes acting on
already formed galaxies.
The alternative picture is that 
morphological properties might be set early on at galaxy formation, 
and are due to the very {\it nature} of the object, such as the mass
of its hosting dark-matter halo.
%
%
A direct systematic investigation of these processes at different cosmic
epochs 
has been so far limited by the relatively small size of the samples
for which HST morphological information is available (e.g. Smith et
al. 2005), and thus by the difficulty to adequately sample the full
range of environments.

%
\begin{figure*}
\epsscale{1.0}
\plotone{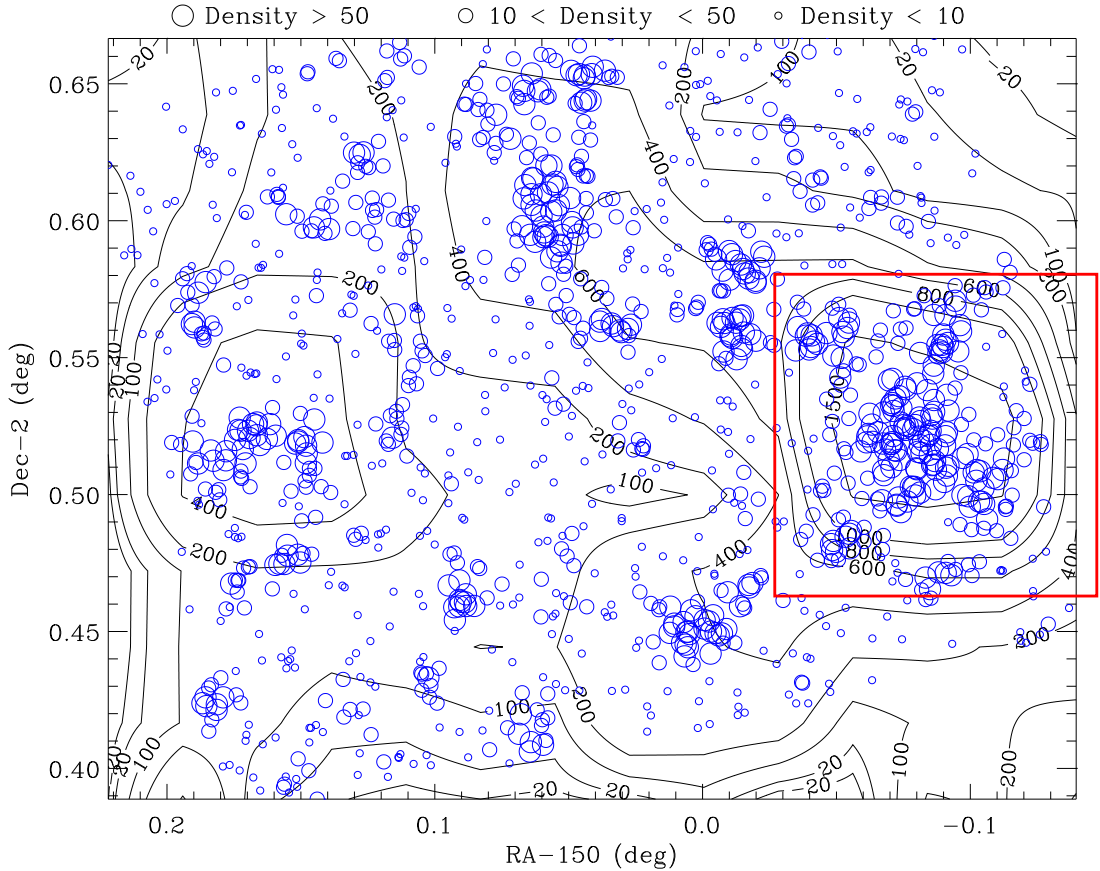}
\caption{Galaxy distribution over the region of the COSMOS field
containing the large-scale structure at $\left<z\right>=0.73$, showing
all galaxies with $M_V<-20.27$ and photometric redshift within 
$\Delta z = \pm 0.12$.   The size of the circles is proportional to
the measured value of the (background corrected) local surface density
$\Sigma_{10}$ at the object's position, while the contours give the
loci of constant value for the same quantity, in units of h$^2$
  Mpc$^{-2}$, after smoothing with a $s\times 3$ boxcar.  As explained
  in the text, $\Sigma_{10}$ is background-subtracted and only
  galaxies over a mosaic of 30 HST tiles are used in this paper (note
  missing galaxies in corners).  Negative contours near the corners
  are an artifact due to this. 
Note how also some galaxies not residing in the highest
large-scale peaks, can experiment relatively high values of
local density.  This picture corresponds at $z=0.73$ to a comoving area of
$ 12.7 \times 9.5$ h$^{-2}$ Mpc$^{2}$.  The box identifies the richer
cluster in the structure.
}
\label{dens_map}
\end{figure*}
%
The Cosmic Evolution Survey (COSMOS), the largest contiguous survey
ever with HST, provides for the first time a combined data set capable
of simultaneously exploring large-scale structure and detailed galaxy
properties (luminosity, size, color, morphology, nuclear activity) out
to a redshift approaching $z=1.5$.  In this paper, we use deep,
panoramic multi-color imaging from SUBARU Suprime-Cam, extended X-ray
imaging from XMM-Newton and the superb resolution HST data to
characterise an outstanding large-scale structure at $\left
<z\right>\simeq 0.73$ within the COSMOS field and explore in detail
the relationship between galaxy morphology and the environment (local
density).  In a second, parallel paper (Cassata et al. 2007), we
instead discuss for the same sample the environmental dependence 
and morphological composition of the galaxy color-magnitude diagram.
Additionally, Capak et al. (2007b) use the full COSMOS
data set to explore the evolution of the MD relation for $z=[0.2,1.0]$ over the
whole range of environments. 

The paper is organised as follows: in \S~2 we provide a brief
description of the COSMOS data sets used in this work; in \S~3 we
discuss the properties of the large-scale structure, in terms of
galaxy surface density, weak lensing projected mass distribution and
X-ray surface brightness; here we study in particular the lensing and
X-ray properties of dominant cluster of galaxies in the structure; in
\S~4 we introduce our estimates of galaxy morphology, presenting the
measurement of the MD relation; in \S~5 we show how galaxy
morphologies are affected by large-scale features detected in the
intra-cluster medium of the dominant cluster; finally, we summarise
our results in \S~6.

We adopt throughout the paper a cosmological model with
$\Omega_M=0.3$, $\Omega_\Lambda = 0.7$ and, in general, parameterise
explicitly the Hubble constant as h=H$_o/100$ when quoting linear quantities,
densities and volumes.  When computing absolute magnitudes and X-ray
luminosities we specialize it to $h=0.7$.

\section{The Data}


The COSMOS project is centred upon a complete survey in the
near-infrared F814W ($\sim I+z$) band using the Advanced Camera for
Surveys (ACS) on board HST of a 1.7 square-degree equatorial field
($\alpha$(J2000) = $10^{\rm h} ~ 00^{\rm m} ~ 28.6^{\rm s}$,
$\delta$(J2000) = $+02^\circ ~ 12' ~ 21.0''$, Scoville et al. 2007).
With an ACS field of view of 203 arcsec on a side, this required a
mosaic of 575 tiles, corresponding to one orbit each, split into 4
exposures of 507 seconds dithered in a 4-point
line pattern.  The final 575 ``MultiDrizzled'' (Koekemoer et
al. 2002) images have absolute astrometric accuracy of better than 0.1
arcsec.  A version with 0.05 arcsecond pixels was used to measure the
morphology of galaxies in the structure; a more finely sampled version
with 0.03 arcsecond pixels was required to measure the shapes of more
distant galaxies behind the cluster, which are needed for the weak
lensing analysis.  More details and a full description of the ACS data
processing are provided in Koekemoer et al. (2007).


The COSMOS field is also the target of further observations in several regions
of the electromagnetic spectrum (see Scoville et al. 2007 for an
overview).   Extensive ground-based imaging has been performed
using Suprime-Cam (Kaifu et al. 2000) on the 8.2~m Subaru Telescope on
Mauna Kea, in six bands, $B_j$, $g^+$, $V_j$, $r^+$, $i^+$, and $z^+$, with a
$5\sigma$ sensititivity in AB magnitudes corresponding to 27.3,
26.6,27.0,26.8, 26.2, 25.2, respectively (Taniguchi et al. 2007, Capak
et al. 2007a).  These data were complemented with further imaging in
$u^\star$ $i^\star$ and $K_s$ from CFHT-Megacam, KPNO and CTIO respectively.
The resulting global photometric catalogue contains around 796,000
objects to $I_{AB}=26$.  More details on the observations and data
reduction are given in Taniguchi et al. (2007), Capak et al. (2007a)
and Mobasher et al. (2007).

Photometric redshifts were derived using a Bayesian Photometric
Redshift method (BPZ) (Benitez 2000; Mobasher et al. 2007), which uses
six basic SED types, together with a loose prior distribution for
galaxy magnitudes.  The output of the code includes, in addition to
the best estimate of the redshift, the 68 and 95\% confidence
intervals, the galaxy SED type and stellar mass and its absolute
magnitude in several bands.  The latter is particularly important for
the current analysis, where the morphology-density relation has been
studied for different luminosity thresholds.  Based on nearly 1200 objects for which
spectroscopic observations are available (not considering catastrophic
failures), the difference between spectroscopic and photometric
redshifts has a typical {\it rms} value $\sigma_z/(1+z)\simeq 0.03$.
This figure is consistent with the restricted statistical sample of
spectroscopic redshifts available within the large-scale structure at
$\left<z\right>=0.73$ studied here (\S~\ref{proj_dens}).  To
further improve on this, for the present analysis we restrict the
photometric redshift sample to $I_{AB}<24$, corresponding roughly, at
the structure mean redshift and for a mean SED type, to an absolute
magnitude $M_V>-18.5$ in the adopted cosmology.  More details on the
quality of the photometric redshifts can be found in Mobasher et
al. (2007).

Relevant for the present work is the extensive X-ray survey that is
nearing completion using the XMM-Newton satellite, for a total granted
time of 1.4 Msec (Hasinger et al. 2007, Cappelluti et al. 2007).  The
mosaiced data used in this paper correspond to the first 0.8 Msec,
providing a median 40 ksec effective depth, taking vignetting into
account.  This in turn results in a typical flux limit for extended
sources of $\sim 2 \times 10^{-15}$ erg s$^{-1}$ cm$^{-2}$, an
unprecedented value for a survey this size.  A first catalogue of
X-ray selected clusters in the COSMOS field is also being presented in
this same volume (Finoguenov et al. 2007).
A full description of the the XMM observations can be found in Hasinger et
al. (2007).  

In this paper we present a pilot study that concentrates on a small
subset of the COSMOS data, covering an area of about $ 22^\prime$ in
Right Ascension and $19^\prime$ in Declination.  This corresponds to
30 ACS tiles and is covered by all the multi-wavelength data
sets described above.  For several of our computations (as e.g. to
estimate the overall mean galaxy background), we shall also make use
of the full COSMOS catalogue over the entire $\sim 2$ square degrees.

\section{A large-scale structure at z=0.73 in the COSMOS field}

A search for large-scale structures within the COSMOS field based on
adaptive-smoothing within photometric redshift ``slices'', finds --
among a number of structures found at different redshifts -- one
conspicuous peak in the redshift interval $z=[0.65,0.85]$ (Scoville et
al. 2007c).  As we shall show in the following sections, the overall
structure is evident in both a weak-lensing analysis of the whole
COSMOS field (Massey et al. 2007) and in the distribution of X-ray
emission from the XMM mosaic (Finoguenov et al. 2007).  Specifically,
the strongest peak in the lensing map is also one of the most
prominent extended X-ray sources in the COSMOS field (Massey et
al. 2007). In the following sections, we shall probe the reliability
of these different observations in locating structures and
characterize the environment and its correlation to galaxy
morphologies.

\subsection{Large-scale structure from galaxy counts}
\label{proj_dens}

%
%
\begin{figure}
\epsscale{1.}
\plotone{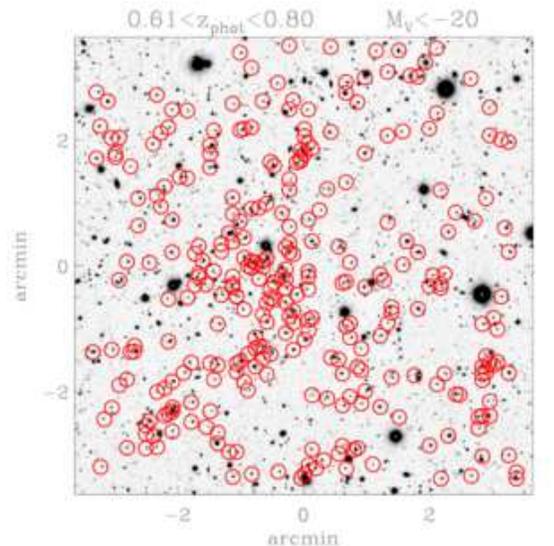}
\caption{A zoom on the cluster of galaxies dominating our
  $\left<z\right>=0.73$  large-scale
structure, corresponding to the box of Figure~\ref{dens_map}. Here a
sub-region of the SUBARU $z$-band 
image is shown, with superimposed the positions of detected galaxies in
the catalogue (open circles).  To enhance the structure definition on
this more restricted area, these correspond to a slightly
broader range in absolute magnitudes than those in Figure~\ref{dens_map}, i.e. 
$M_V<-20$.
Note the ``filament'' of galaxies running from S-E to N-W (with a
``wiggle'' corresponding to the central cluster), with galaxy
concentrations corresponding to X-ray blobs in the X-ray map
(Figure~\ref{xray_overlay}). }
\label{photoz_overlay}
\end{figure}
%
Scoville et al. (2007c) use an adaptive kernel technique to map
the surface density of galaxies in intervals of photometric redshift.
Here, we aim at using an estimate of local density that would allow us
to study, in particular, the morphology-density relation within the
structure and compare that consistently with previous measurements. 
Following Dressler (1980) and similar works in the literature, we have
thus computed
the $\Sigma_{10}$ estimator
\begin{equation}
	\Sigma_{10}= {\frac{10} {\pi d_{10}^2}} \,\,\,\,\  ,
\end{equation}
where $d_{10}$ is the projected distance in h$^{-1}$ Mpc (computed
from the angular separation using the galaxy corresponding angular
diameter distance) to the 10-th nearest neighbour.  This is an
adaptive estimate of density, with the major advantage over a fixed
aperture measurement of avoiding being dominated by shot noise in
low-density regions.

Dressler's original work was limited to fairly luminous galaxies
within rich Abell clusters, i.e. well-defined, local structures of
high contrast where the two-dimensional over-density signal produced
by cluster galaxies strongly dominates over the background.  The
estimator $\Sigma_{10}$ is thus in this case a fairly direct
(projected) probe of the 3D density around the selected galaxy, with
negligible contribution from the background/foreground objects.  When
the same definition is applied to the general ``field'', it is
necessary to specify a typical depth over which the integration is
performed, as to remain a meaningful proxy for the true 3D
environment.  A sensible choice is a size of the order of the galaxy
correlation length or the typical group/cluster velocity dispersion
($r_o \sim \sigma_v/H_o \sim 5 - 10 \hmpc$).  The ten nearest
neighbours around each galaxy are then counted within such a redshift
cylinder and used to estimate $\Sigma_{10}$.  This kind of approach
has been used at high redshift by Smith et al. (2005), who extend
their study also out of rich clusters. The problem is further exacerbated
when only photometric redshifts are available.  With
typical errors of the order of $0.05(1+z)$ in redshift (corresponding to
moving a galaxy along the line of sight by $\sim 100
\hmpc$ or more), the choice of the redshift interval over which
integrating galaxy counts becomes crucial.  This is essentially a
compromise between a range large enough to recover all true
companions scattered along the line of sight by the redshifts errors
(completeness) and small enough to minimize fake
projections (contamination).  The situation in our case, with a sample
centred on a well-defined main structure located at $z\sim 0.73$, is
slightly simplified with respect to a more general case of estimating
local density for galaxies within a generic redshift slice (Cooper et
al. 2005).  Following similar work
in the literature (e.g. Kodama et al. 2001; Smith et al. 2005; Postman
et al. 2005), we looked first at the dispersion in the differences
between spectroscopic and photometric redshifts.
The overall standard deviation in the COSMOS photometric redshift
catalogue, based on the larger set of $\sim 1000$ spectroscopic
redshifts from the first runs of the zCOSMOS redshift survey (not
covering this area yet) is $\sigma_z\simeq 0.03(1+z)$ (Mobasher et
al. 2007, Lilly et al. 2007).  In addition, VLT-FORS1 spectroscopy of
15 red galaxies around the central cluster of this structure
(Comastri, private comm.), give a comparable value for this redshift, 
$\sigma_z=0.056$, with a median redshift for the cluster of
$z=0.7318$.  In the following, we shall assume that the bulk of our
structure within the $22^\prime\times 19^\prime$ area under study,
lies at the mean redshift $\left<z\right>=0.73$.  This assumption is
fully justified by the distribution of photometric redshifts and 
allows us to simplify our computations.

On the basis of these indications, we use the simple approach of
selecting a single slice centered at z=0.73 (similar to Kodama et
al. 2001) and with appropriate thickness, and then correct
statistically the measured $\Sigma_{10}$ of each galaxy for the
background contamination.  We obtain a first estimate the robustness
of the method and of the most appropriate thickness to be chosen for
the redshift slice, by exploring three different sizes, $\delta z =
\pm 0.06$, 0.12, 0.18 corresponding to $\pm 1$, 2 and $3\sigma_z$'s
(note that these values comfortably include the velocity dispersion of
any known system, being this about 20 times smaller than the redshift
errors).  Additionally, in Appendix A we present a more direct test of
how well we are able to recover the true density, given the
photometric redshift errors, based on the analysis of realistic mock
samples of the COSMOS survey constructed from the Millennium
Simulation (Kitzbichler \& White 2006).  A more sophisticated approach
to measure local densities from the photometric sample has been
adopted in our parallel work on the evolution of the
morphology--density relation from COSMOS (Capak et al. 2007b); in that
paper, we use a moving $\pm \delta z$ cylinder centred at the redshift
of each galaxy (as in Smith et al., 2005), and describe each galaxy
not as a Dirac delta function in redshift space, but as a broader
probability distribution corresponding to its specific photometric
redshift likelihood function.  As shown by our
tests, the simpler approach adopted here is more than adequate for the
specific geometry of our sample, producing results fully consistent
with Capak et al. (2007b).

One further source of uncertainty is related to the background
correction to the measured density.  In principle, with a sufficiently
thin slice and spectroscopic redshifts, one might reasonably assume
that, outside filaments and clusters, the density of galaxies is
practically zero, i.e. that no background correction is needed.  
In the case of photometric redshifts, however, where a much
thicker slice has to be used,  a number of background and foreground
galaxies not belonging to the inner structure will be included, 
shifting the background surface density to a non-zero
value.   We assume that this term is dominant and estimate the
background correction as the median value of $\Sigma_{10}$ over the
full COSMOS 2 square degrees, in the corresponding redshift slice.  As
shown in the Appendix, this correction works very well and corrects
for the bias in $\Sigma_{10}$ that would tend to systematically
overestimate local densities. 
\begin{figure}
\epsscale{1.}
\plotone{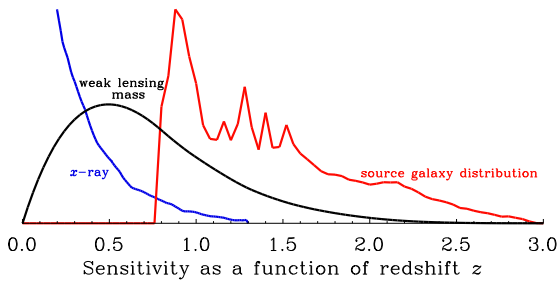}
\caption{Relative sensitivity of the lensing and X-ray analyses to a
  fixed mass as a function of redshift, with arbitrary
  normalization. Also shown is the redshift distribution of galaxies
  from which shears were measured, throughout the COSMOS field. All of
  the peaks below $z=1.2$ (including that at $z=0.73$), correspond to
  real structures; the field is small enough to be subject to sample
  variance.  Above that, photometric redshift degeneracies due to the
  finite number of available colors cause estimates to accumulate at
  discrete redshifts.}
\label{lensing-sensitivity-function}
\end{figure}
%
\begin{figure*}
\epsscale{1.0}
\plotone{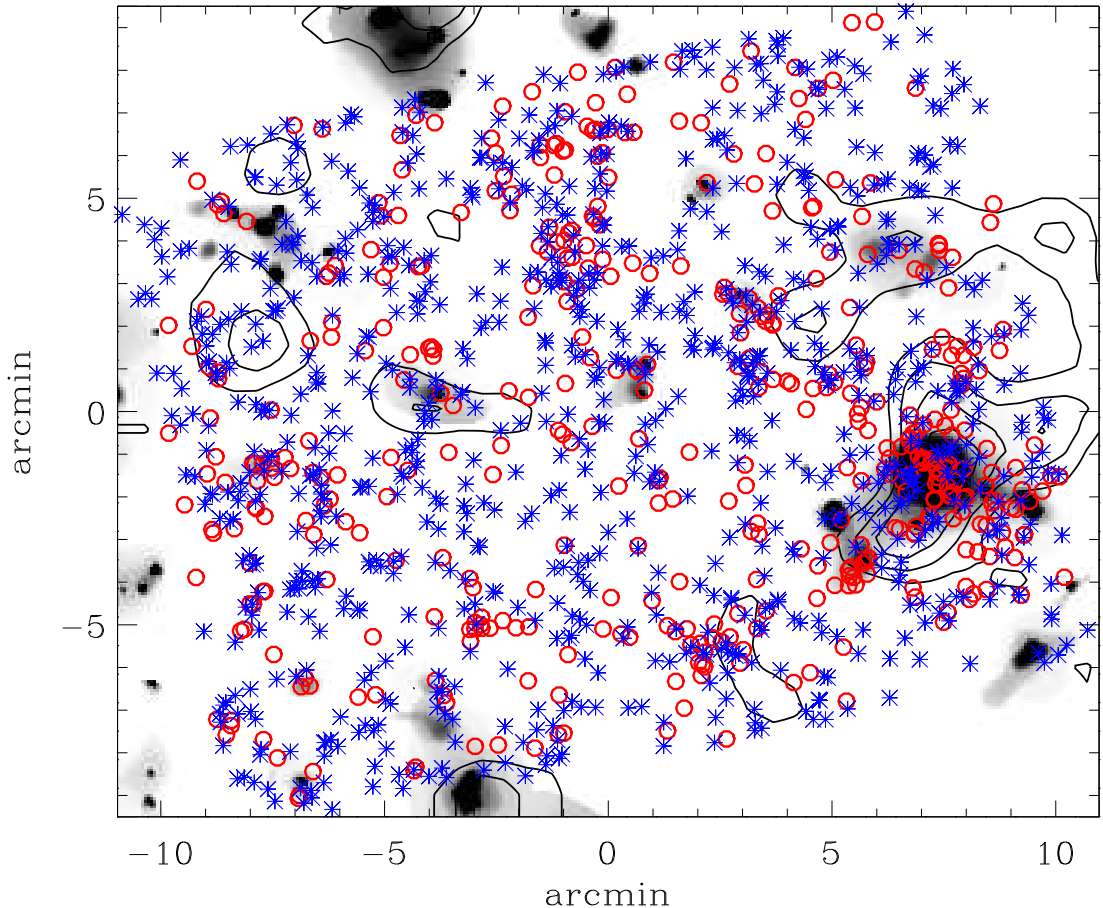}
\caption{Overview of the large-scale structure at
  $\left<z\right>=0.73$, as seen in the SNR map of the weak-lensing
  projected mass 
  reconstruction (contours, Massey et al. 2007), X-ray surface
  brightness (gray scale background, Finoguenov et al. 2007), and 
  distribution of galaxy morphological types.  Weak-lensing contours
  describe positive $2\sigma$, $3\sigma$, $4\sigma$ and $5\sigma$
  levels of signal to noise in the convergence field $\kappa$ of the
  lens (cf. Hamana, Takada \& Yoshida 2004), while the X-ray
  grey-scale levels starts at $1\sigma$ over the background and are
  spaced by $1\sigma$.  Only galaxies brighter than Dressler's limit
  ($M_V<-20.27$ at this redshift, for h=0.7) are plotted, with
  early-type galaxies indicated by red circles and late-type galaxies
  by blue asterisks.  Contour and grey-scale levels are spaced in
  terms of $rms$ values over the background.  }
\label{morph_map_bright}
\end{figure*}
%

Finally, another choice needed to make $\Sigma_{10}$ fully consistent
with previous works, is that of the range of absolute magnitudes to
include in the computation.  In the original Dressler (1980) paper,
the limit was set to $M_V=-20.4$, using $H_o=50$.
Considering a brightening of the luminosity function of 0.6 mag to
$z=0.73$ (e.g. Zucca et al. 2006), which is comparable to values
adopted in Smith et al. (2005) or Treu et al.  (2003), and converting
to $h=0.7$, 
this corresponds to a value $M_V<-20.27$ for our data.
We note that there is little
agreement in the literature on the appropriate amount of brightening to
be applied.  Part of the difficulty arises because the actual average
brightening depends significantly on the relative contributions of 
different spectral types: the luminosity function of late-type
galaxies evolves more rapidly than that of
early types, as shown by the VVDS survey results (Zucca et al. 2006).
As we shall show in \S~\ref{morph}, the resulting MD relation does
not depend significantly on either the choice of the redshift interval
or a brightening between 0 and 1 magnitudes.

In Figure~\ref{dens_map}, we plot contours of constant projected
density $\Sigma_{10}$, as measured from the $0.61<z_{phot}<0.85$ slice.
Luminous galaxies ($M_V<-20.27$) with photometric redshifts in the
same interval are indicated by circles, with symbol sizes proportional
to the value of local density.  We remind that this analysis uses only
30 ACS tiles, out of the whole 575 COSMOS-ACS images.  These tiles
cover an inclined rectangle over this area, as indicated by the lack
of objects in the corners of the figure. At least six galaxy
concentrations of different strength are evident in this figure, some
of which aligned along filaments.  A zoom over the most prominent
density peak
is displayed in Figure ~\ref{photoz_overlay}.  Galaxy positions for
the luminous galaxies are over-plotted on a section of the $z^+$-band
Subaru image.
Note the ``S''-shaped filament of galaxies seemingly running
SE - NW, with a sharp bend near the center.   


\subsection{Large-scale structure from weak lensing}

%
\begin{figure}
\epsscale{1}
\plotone{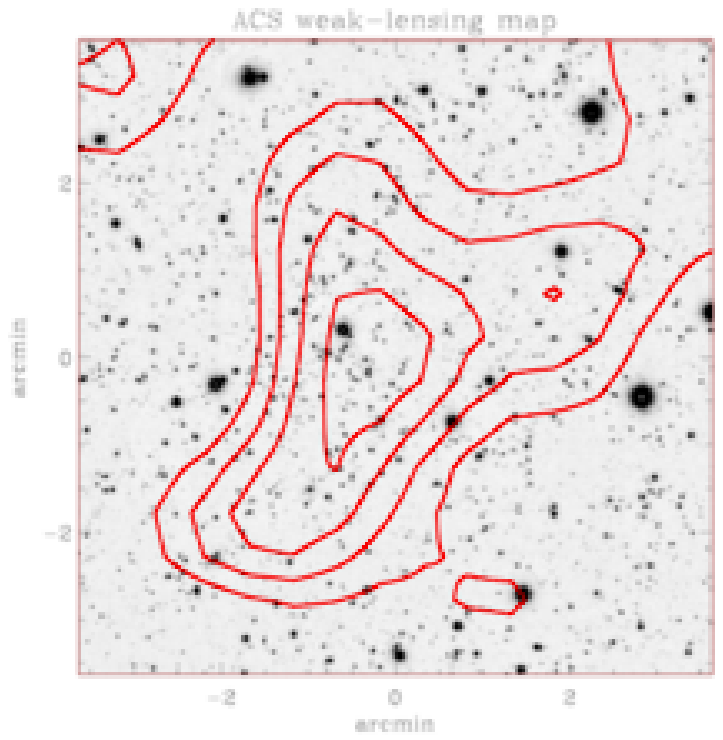}
\caption{Zoom of the mass reconstruction contours 
from the E-modes of the weak lensing
shear, 
over the same area shown in
Figure~\ref{photoz_overlay}.
Again, the contours show 2-, 3-, 4- and $5\sigma$ levels of
signal to noise in the convergence field. None of these are visible in 
the $B$-mode signal in this region. If all of the
mass lies in a single redshift plane, convergence is proportional to
mass. Note the comments in the text about projection effects and the
sensitivity of lensing to a broad range of redshifts, but not the same
redshifts as optical or x-ray data. The image underneath is again 
the SUBARU z-band image.
}
\label{lensing_overlay}
\end{figure}


The observed shapes of background galaxies are slightly distorted as
light from them is deflected around the gravitational field of
foreground structures. The high resolution HST coverage of the COSMOS
survey allows us to measure these distortions and to reconstruct the
weak lensing ``shear field'' with unprecendented precision. To
reconstruct the distribution of mass in this field, we first measured
the shapes of galaxies using the ``RRG'' algorithm by Rhodes,
Refregier \& Groth (2000). The specific application of this method to
the COSMOS data is fully described in Leauthaud et al.\ (2007) and
Massey et al.\ (2007). The galaxy shapes were then corrected for the
effects of instrumental Point-Spread Function (PSF) convolution using
the PSF models described in Rhodes et al.\ (2007).

Photometric redshifts from Mobasher et al.\ (2007) were used to
identify 37.6 galaxies per square arcminute behind the cluster, with a
mean (median) photometric redshift of 1.50 (1.25). Reaching this
number density for such a high-redshift structure required a slight
reduction in the size and magnitude thresholds used for the analysis
of the entire COSMOS field by Massey et al.\ (2007). This is possible
with a minimum impact in signal to noise because of the
strong signal in this region. However, we have increased the estimate
of noise due to possible shear calibration error to $12\%$ to include
the additional uncertainty in the shapes of very faint galaxies.  The
shear calibration error represents uncertainty in a shear measurement
method's ability to deconvolve the shapes of faint galaxies from the
instrumental PSF and to accurately measure their
ellipticities. The accuracy of our shear measurement method has been
estimated using simulated images that resemble COSMOS data but which
contain a known shear signal. In fact, as popularised by the Shear
TEsting Programme (STEP; Heymans et al. 2006; Massey et al. 2007b),
shear measurement errors can actually be parameterized as two numbers:
the shear calibration error $m$ (multiplicative error), and the
residual shear offset $c$ (additive error), although in the cluster
lensing regime, where the signal is large, the latter source of error
is negligible. Both are estimated and tabulated in the parallel paper
by Leauthaud et al. (2007).

The observed shear field from the entire COSMOS field was transformed
into a convergence $\kappa$ map using the method of Kaiser \& Squires
(1993), filtered by a Gaussian of rms width $100\arcsec$. The
convergence field is related to the Newtonian gravitational potential
$\Phi$ as

\begin{equation}
\kappa ~ \equiv ~ \frac{1}{2} \left(\frac{\partial^2\Psi}{\partial x^2} + \frac{\partial^2\Psi}{\partial y^2}\right) ~,
\label{theory:eq:convergencedefn}
\end{equation}

\noindent where

\begin{equation}
\label{eq:psi_dchi}
\Psi_{ij} = \int g(z) \partial_{i} \partial_{j} \Phi ~{\rm d}z ~,
\end{equation}

\noindent $\partial_{i}$ is the comoving derivative perpendicular
to the line of sight and $g(z)$ is the radial sensitivity function 

\begin{equation}
g(z) \equiv 2 \int_{z}^{\infty} \eta(z)~
   \frac{D_A(z)D_A(z^\prime-z)}{D_A(z^\prime)}~a^{-1}(z)~{\rm
     d}z^\prime\,\,\, .
\label{eq:gz}
\end{equation}

\noindent In this expression, $a(z)$ is the cosmological scale factor,
$D_A$'s are angular diameter distances and $\eta(z)$ is
the distribution of background source galaxies, normalised such that

\begin{equation}
\int_0^{\chi_{h}} \eta(\chi) ~{\rm d}\chi = 1 ~.
\end{equation}

\noindent As shown in Figure~\ref{lensing-sensitivity-function}, in
our analysis $g(z)$ peaks between 
$z=0.3$ and $z=0.8$. Lensing is
less sensitive to structure closer or more distant than this.

In Figure~\ref{morph_map_bright}, we have plotted the contours of
constant SNR in the reconstructed projected mass distribution over our
study field (see \S~\ref{structure_overview} for more discussion and
comparison to the galaxy and X-ray distributions).
Figure~\ref{lensing_overlay}, instead, shows a zoom 
on the strongest peak, centred at RA=149.922, DEC=2.515, and coinciding
with the dominant cluster of galaxies within the structure already shown in
Figure~\ref{photoz_overlay} (cf. box in Figure~\ref{dens_map}).  

It is of interest to estimate the mass of this cluster. To this end,
we first need to fix the geometry of the lens. We assume that all of
the foreground mass lies at $z=0.73$ and that the background galaxies
lie at exactly the redshifts indicated by their photo-$z$
estimates. We
then first apply the mass aperture statistic $M_{\mathrm ap}$ of
Schneider et al. (1998)

\begin{equation}
M_{\rm ap}(\theta) \equiv \int_\theta
\kappa\left(\vec\vartheta\right)~U\left(|\vec\vartheta|;~\theta\right)~{\rm d}^2\vec\vartheta ~,
\label{eqn:_01}
\end{equation}

\noindent 
which calculates the value of the surface mass inside the radius
$\theta$ on the sky, convolved with a compensated filter

\begin{equation}
U(\vartheta;~\theta) \equiv \frac{9}{\pi\theta^2}
\left(1-x^2\right)\left(\frac{1}{3}-x^2\right) H(\theta-\vartheta) ~,
\label{eqn:_02}
\end{equation}

\noindent where 
$x=\vartheta/\theta$ and $H$ is the Heaviside step function (see Clowe
et al. 2006, for a similar application).  Using a $\theta=100\arcsec$
window, we find 
\begin{eqnarray} \label{eqn:maperrors}
M_{\mathrm ap}(100\arcsec)
&=& (1.10 \pm 0.16~^{+0.09}_{-0.13} \pm 0.13) \times10^{14} {\rm M_\sun}\\
&=& (1.10~^{+0.22}_{-0.24}) \times10^{14} {\rm M_\sun}~,
\label{eqn:map}
\end{eqnarray}
%
\noindent where the three sets of $1\sigma$ error bars in
equation~(\ref{eqn:maperrors}) respectively correspond to statistical noise in
galaxy shape measurement, including their intrinsic ellipticities; background
and foreground redshift uncertainty; and shear calibration uncertainty. These
have been combined in quadrature in equation~(\ref{eqn:map}).  

The $M_{\mathrm ap}$ statistic allows us to assess the significance of
our measurements: gravitational lensing by a single cluster produces
a shear field that is aligned tangentially around the mass overdensity
(a void would produce a radial shear field). An independent component
of the shear field can be separated: by analogy with electromagnetism,
this is known as a ``curl'' or $B$-mode, and can be measured by
rotating galaxies by $45^\circ$ in advance. Since it is unphysical, it
is expected to be zero in the absence of systematics, and provides a
good estimate of both systematics plus statistical errors. Around this
cluster, we measure the $B$-mode equivalent of $M_{\mathrm ap}$ to be
\begin{eqnarray}
M_\perp(100\arcsec)
&=& (1.99 \pm 1.55~^{+0.15}_{-0.23} \pm 0.24) \times10^{13}  {\rm M_\sun}\\
&=& (1.99~^{+1.58}_{-1.59}) \times10^{13}  {\rm M_\sun}~,
\end{eqnarray}
%
\noindent where the error bars are the same as before. That this is an order of
magnitude below the mass signal confirms that systematic effects have been
successfully purged from the lensing analysis.

\begin{figure}
\epsscale{1.}
\plotone{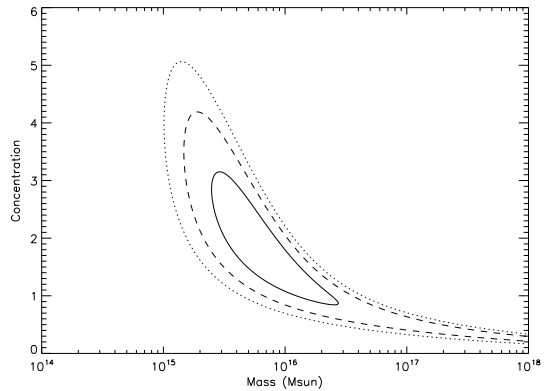}
\caption{Constraints on the total mass and the Concentration parameter
  of the extended dark-matter halo in the dominant cluster, obtained
  by fitting a circular NFW profile to the weak lensing shear in
  radial bins from the center.  The solid, dashed and dotted lines
  show 68\%, 95\% and 99\% confidence limits respectively. Note that
  there is much more mass at this location than that suggested by
  either the $X$-ray peaks or the lensing aperture mass in a corresponding
  region.}
\label{NFW-fit-to-shear}
\end{figure}

However, as evident from Fig.~\ref{dens_map} the cluster is in a complex
large-scale environment, shows significant substructure in the process
of merging, and is at the nexus of several filaments (see also
Scoville et al. 2007, Massey et al. in preparation). To estimate the
total mass from its extended dark matter halo, we have then fitted a
circularly symmetric NFW profile (Navarro, Frenk \& White 1997) to the
radial shear field, using the technique of King \& Schneider
(2001). As shown in figure~\ref{NFW-fit-to-shear}, this finds a much
larger mass. The best-fitting model has
$M_{NFW}=(6\pm3)\times10^{15}M_\sun$.
As we shall show in the next section, this cluster 
is also one of the most prominent extended X-ray sources in the COSMOS
field.


\subsection{Large-scale structure from X-rays}
\label{X-ray}
%
\begin{figure}
\epsscale{1.}
\plotone{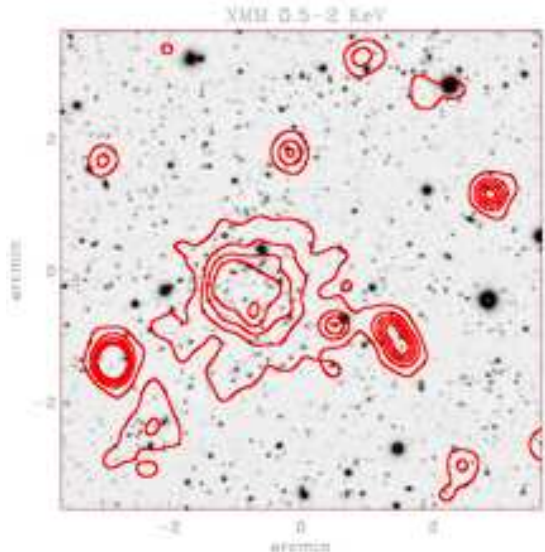}
\caption{ Contours of constant X-ray surface brightness in the
  [0.5-2.0] keV band from XMM 
(smoothed with a 20 arcsec boxcar filter), corresponding to
1-,2-,3-,4- and 5-$\sigma$ levels above the background, superimposed
on the SUBARU $z$-band image.  
The X-ray data are a combination of the
background-corrected/flat-fielded MOS and PN images (see Finoguenov et
al. 2007). The area is the same as in
Figure~\ref{photoz_overlay}.  Note the large
extended source in the center, corresponding to the strongest
concentration of galaxies in the photometric redshift slice and the
strongest peak in the lensing convergence maps.
Despite the different resolution in the lensing and X-ray maps, the
match between these two and with the galaxy distribution is
remarkable.  }
\label{xray_overlay} 
\end{figure}
The grey-scale levels in Figure~\ref{morph_map_bright} shows the
distribution of X-ray surface brightness (in units of standard
deviations from the background) over the area under study, obtained
from the reduced XMM mosaic described in Hasinger et al. (2007) and
Finoguenov et al. (2007).  Figure~\ref{xray_overlay} shows a more
detailed zoom over the same cluster sub-area of
Figures~\ref{photoz_overlay} and \ref{lensing_overlay}.  The galaxy
cluster emerges as a powerful extended X-ray source, with an observed
flux in the $[0.5-2.0]$ keV band of $4.56 \pm 0.13 \times 10^{-14}$
erg cm$^{-2}$ s$^{-1}$, corresponding at $z=0.73$ to an X-ray
luminosity (in the $[0.1-2.4]$ keV band) of $1.56\pm 0.04 \times
10^{44}$ erg s$^{-1}$ (Finoguenov et al.\ 2007).  

In addition to the
central cluster, a number of fainter sources are detected.  Comparison
to Figure~\ref{photoz_overlay} clearly shows some of them obviously
coinciding with galaxy concentrations.  Despite the different
selection functions, the structure seen in
the galaxy distribution and in the lensing and X-ray maps is
remarkably similar.  Note in particular the extension toward
South-East in the lensing map, coinciding with the secondary galaxy
condensation along the ``S''-shaped filament and with an extended
source in X-rays.

%
\begin{figure}
\epsscale{1.0}
\plotone{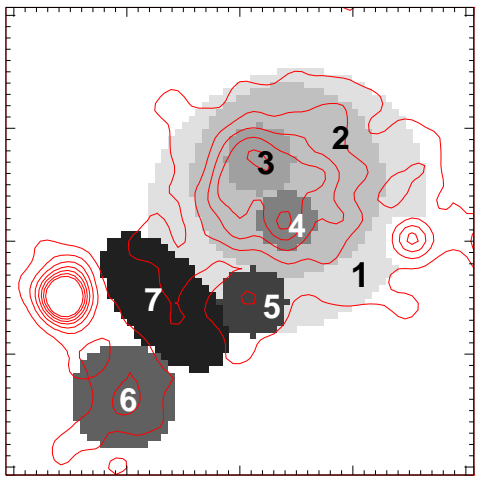}
\caption{Schematic map of the 7 regions, described in
  Table~\ref{table:tx}, into which the X-ray emission from the cluster
  of Figure~\ref{xray_overlay} has been sub-divided, to estimate the
  distribution of X-ray temperatures.  The contours of X-ray surface
  brigthness are repeated here to ease comparison to previous figures.
}
\label{tx_regions} 
\end{figure}
%
\begin{table}[h]
\begin{center}
\caption{Cluster X-ray temperature measurements
\label{table:tx}
}
\vspace{3mm}
\begin{tabular}{cll}
\tableline
zone & $kT ({\rm keV})$ & Notes  \\ 
\tableline
1 &	$2.39_{-0.49}^{+0.95}$ & cluster outskirts \\
2  &	$3.51_{-0.46}^{ +0.60}$ & cluster global ICM \\
3 & 	$10.30 _{-3.18}^{ +6.14}$  & possible shock region / point source?\\ 
4 & 	$3.49_{-0.63}^{ +0.96}$  & cluster core \\
5 &  	$2.39_{-0.37}^{ +0.64}$ & sub-structure\\
6 & 	$2.50_{-0.51}^{ +0.80}$ & ``infalling'' group\\
7 & 	$2.85_{-0.80}^{ +1.02}$  & intermediate region 6 - 1 \\ 

\tableline
\end{tabular}
\end{center}
\end{table}

Combining these complementary data sets allows us to start building up
a coherent picture of large-scale structure in this region, which
appears to be dynamically very active.  For example, looking at the
X-ray map of Figure~\ref{xray_overlay}, one notices a clear asymmetry
in the surface brightness distribution of the central cluster, which
also shows some significant (at XMM resolution)
substructure.  The main feature is a significantly steeper
gradient in surface brightness toward North-East, with respect to a
looser profile in the opposite direction, possibly suggesting the
presence of a shock front.  This edge coincides also with the sharp
bend visible in the ``S''-shaped galaxy filament of
Figure~\ref{photoz_overlay}, another indication that it might
correspond to the main interface of an ongoing major merging event
between two sub-structures along the SW-NE direction.  

The depth of the XMM observations
allows us to look for the fingerprint of a shock, by
studying in some detail the X-ray temperature map over the cluster
area, and in particular near the NE rim.  
We have thus divided the area into seven regions, estimating
for each of them the X-ray temperature.  This has been performed by fitting an
emission spectrum from collisionally-ionized diffuse gas, computed
using the APEC code (v1.3.1., see http://hea-www.harvard.edu/APEC/ for
more information).    The
location and numbering of each region is reported in the sketchy
map of Figure~\ref{tx_regions}, and the corresponding meaning and
resulting temperature measurements are reported in
Table~\ref{table:tx}.   

The cluster ICM is found to have a global X-ray temperature
(integrated over a radius $r_{500} = 1.4$ arcmin, corresponding to
0.43 h$^{-1}$ Mpc) of $kT_X = 3.51_{-0.46}^{ +0.60}$ keV.  Using the
relation derived by Pacaud (private comm.) from the data of Finoguenov
et al. (2001)
\begin{equation}
M_{500}=2.36\times10^{13}M_\odot \times
T_X^{1.89} E(z)^{-1} \,\,\, ,
\end{equation}
where $E(z)$ is the usual cosmological function specialized to our
cosmology, we obtain a mass $M_{500}=1.69_{-0.38}^{+0.58} \times
10^{14}M_\odot$.   Alternatively, using the $M-T_X$ relation from Vikhlinin et
al. 2006, and accounting for evolution in the scaling relations as
in Kotov \& Vikhlinin (2005), we obtain $M_{500}=1.57_{-0.30}^{+0.43}
\times 10^{14}M_\odot$.  The difference between these two estimates is
well within the intrinsic uncertainties.

Most interestingly, the measurements show a temperature jump
corresponding to the ``edge'' seen in the surface brightness (region
3): in this region we have $kT_X = 10.30 _{-3.18}^{ +6.14}$ keV.  This
is a strong indication of the presence of a shock front in this
area\footnote{Note that given the strongly non-Gaussian form of the
  probability distribution around this value, this formally large
  $1\sigma$ error -- due to the limited number of counts available
  from such a restricted area -- corresponds in fact to a significant
  deviation ($>99.73\%$ confidence, i.e. $>4\sigma$) of the value of
  $T_X$ with respect to the overall cluster ICM}, as speculated from
the X-ray surface brightness alone. 
A residual uncertainty on this interpretation is related to the
tentative presence of a point-source within region 3, which if
confirmed would pollute our spectral estimate of the temperature.
However, an hardness-ratio map also shows that the NE side of the
cluster core is systematically hotter than the SW one, on a scale
significantly larger than the size of a point source.  Finally, we
note that the observed factor of $1.5\pm0.3$ overestimate in our mass
measurements based on X-ray scaling relations is consistent with being
a consequence of the hotter core of the cluster (that moves the
cluster off the scaling relation), caused by the observed merging
event.  In the not too distant future, it will be possible to study
this system with much more spatial detail, benefiting of the
forthcoming Chandra observations of the COSMOS field.

\section{Galaxy morphology vs. environment at $z\sim 0.7$}
\label{structure_overview}
Having characterised in some detail the large-scale distribution of
galaxies and mass within this particular sub-region at $z\simeq 0.73$,
we turn now to exploring the connection of galaxy morphologies to the
large-scale environment.  Figure~\ref{morph_map_bright} summarises and
compares the information gathered so far.
As already mentioned, this figure shows a remarkable coincidence of
the contours in the surface mass (convergenge) map from lensing with
the diffuse X-ray emission.  This is even more impressive if we
consider that the lensing reconstruction shows mass integrated over a
broad redshift range, with a different sensitivity function to the
$X$-ray and galaxy density analyses (see
Figure~\ref{lensing-sensitivity-function}).  
Due to projection, therefore, one expects some real structure in the
lensing map that does not appear in the other analyses, and vice
versa.  This makes the coincidence of the strongest galaxy
concentrations with the peaks in the X-ray emission and in the lensing
mass distribution even more fascinating.  In addition to the
structures within our study field, note also the two strong
X-ray/lensing blobs partly entering at the top and bottom of the
figure, which are found to be foreground clusters, with mean
photometric redshift of 0.35 and 0.22 respectively (Finoguenov et
al. 2007).

Finally, in this figure we have also indicated galaxy morphological
types, estimated from the ACS data as detailed in the following
section. The distribution of early-type (red circles) and late-type
galaxies (blue asterisks) provides an immediate qualitative evidence
for a significant morphology-density relation within this structure.

\subsection{The morphology-density relation at $\left<z\right>=0.73$}

\label{morph}

In the parallel paper by Cassata et al. (2007), we have measured
automatically non-parametric galaxy morphologies for all galaxies in
the COSMOS field.  In brief, morphologies have been characterized on
the basis of the so-called Concentration, Asymmetry, Clumpiness, M20
and Gini Index (see Cassata et al. 2007 for details and references)
and the sample has been split into two broad morphological classes:
early- and late-type galaxies. Cassata et al. (2005, 2007) show that
it is possible to define a precise region in the space of these 5
parameters, to perform this accurately, with high completeness and
little contamination with respect to a visual classification.
Stars were removed using 
SExtractor according to their CLASS-STAR and FLUX-BEST parameters
re-computed on the ACS data, where this method proves to be reliable,
much more than on ground-based images.

%
\begin{figure}
\epsscale{1.}
\plotone{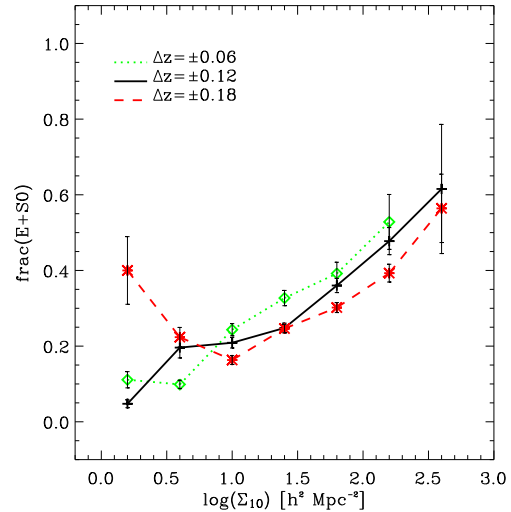}
\caption{The morphology-density relation at $\left<z\right>= 0.73$
as measured over the whole area of Figure~\ref{morph_map_bright} and
for three different thicknesses of the redshift slice.  
The curves give the fraction of early-type galaxies as a
function of local density $\Sigma_{10}$, including only galaxies
brighter than $M_V=-20.27$ as to correspond the value adopted by
Dressler (1980) at $z=0$.
The consistency of the three estimates indicates that the recovered
relationship is robust with respect to the choice of the slice.} 
\label{morph-dens}
\end{figure}
\begin{figure}
\epsscale{1.}
\plotone{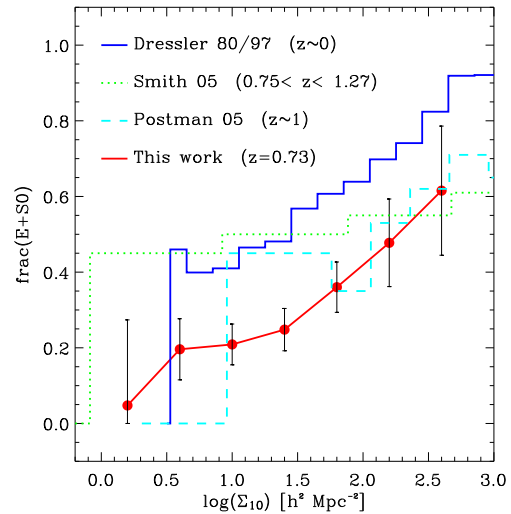}
\caption{Comparison of our estimate of the morphology-density
relation at $\left<z\right>=0.73$ (filled points), with previous measurements at
different redshifts.  
Error bars include also the scatter among the three test slices, as an
attempt to account for systematic uncertainties due to the use of
photometric redshifts.  {\it A posteriori} comparison to the results of the
simulations of Appendix A indicate that they are realistic, and that
the error on the galaxy fraction is dominated by our uncertainty on
the measured density below $log(\Sigma_{10}) \sim 1.5$ and by the
number of galaxies in each bin above this density.}
\label{morph-dens-comparison}
\end{figure}
%
Using this combined data set, we shall explore here in detail the
behaviour of the MD relation within the redshift slice including this
large-scale structure.  With only few spectroscopic redshifts
available over the full area, working on a structure well-confined in
redshift space (while still spanning a significant range of
over-densities) allows us to be reasonably confident that redshift
errors have a negligible impact on our conclusions.  In addition to
the experiments with mock samples discussed in the Appendix, we will
test the robustness of our conclusions explicitly by varying the size
of the sample and its limiting luminosity.  In this analysis we shall
concentrate on the overall MD relation observed at this redshift, its
dependence on galaxy luminosity and in particular to its possible
relation with large-scale structure as described by X-ray emission.
This study is complemented by the work of Cassata et al. (2007), who
explore in detail the morphological composition and environmental
dependence of the color-magnitude relation for galaxies within this
same structure, and that of Capak et al. (2007b), who investigate the
global evolution of the MD relation out to $z\sim 1.2$ using the whole
COSMOS field data.

%
\begin{figure*}
\epsscale{1.0}
\plotone{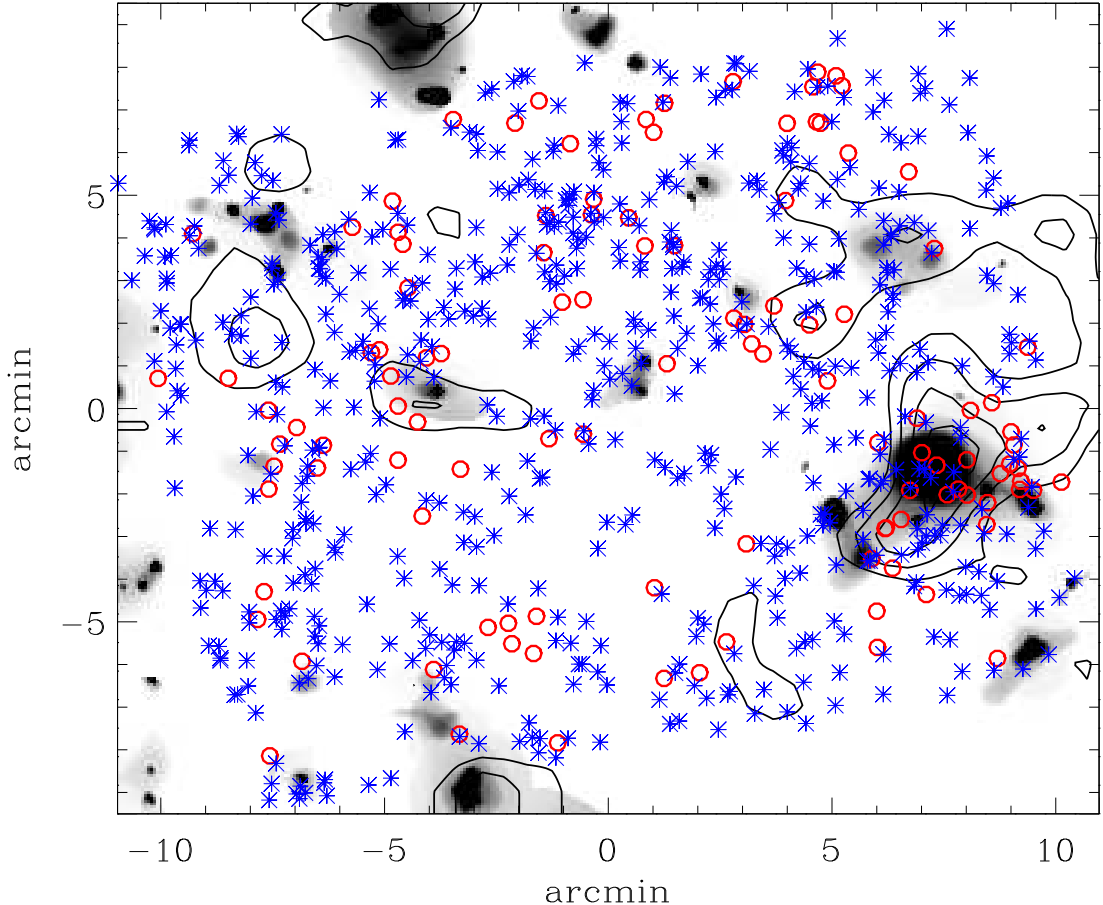}
\caption{Same as Figure~\ref{morph_map_bright}, but now comparing  
the galaxy distribution of galaxies {\it fainter} than Dressler's limit
(i.e. $-20.27<M_V<-18.5$) to the X-ray surface brightness (grey-scale, Finoguenov
et al. 2007) and the weak-lensing projected mass (contours, Massey et
al. 2007) distributions.  
The number of blue star-forming galaxies in this faint sample and
their much smoother surface distribution is evident, especially when
compared to the dominating, highly-clustered population of luminous spheroidals
in  Figure~\ref{morph_map_bright}.  
} 
\label{morph_map_faint}
\end{figure*}
%
The MD relation, i.e. the fraction of early-type galaxies as a
function of $\Sigma_{10}$ for our $\left<z\right>=0.73$ sample,
computed within each of the three redshift slices defined in
\S~\ref{proj_dens} and for $M_V<-20.27$ is shown in
Figure~\ref{morph-dens}.  As discussed, the surface density
$\Sigma_{10}$ has been normalised within the three different redshift
slices, by subtracting the median background value estimated over the
whole COSMOS 2-square-degrees field.  Error bars are here simply the
Poisson's errors on the number of objects in each bin.  However, the
scatter among the three estimates is comparable to the the variance
one obtains by propagating the ensemble errors obtained from the
simulations in the Appendix.  These errors on the derived density can
move galaxies horizontally among the bins, but have a significant
effect only at the lowest densities ($\Sigma_{10}<10$ gal h$^2$
Mpc$^{-2}$).  Note also that the background correction for the three
slices ranges between 20 and 40 h$^2$ Mpc$^{-2}$, and it is shown to
be crucial to reduce the bias introduced by the photo-z errors below
$log(\Sigma_{10}) \sim 1.5$
where the uncertainties become dominant.  Overall, however, the
result is remarkably stable, with the three curves being very
consistent with each other.  This reassures us that the measurement is
not strongly dependent on the size of the redshift slice chosen,
neither does it depend on the details of the background correction.
In the following, therefore, we shall use only the intermediate-size
slice ($\Delta z = \pm 0.12$) as our reference sample.  The
corresponding background correction in this case is of 28 galaxies
h$^2$ Mpc$^{-2}$ (at $M_V<-20.27$).

In Figure~\ref{morph-dens-comparison} we compare our measurement at
$\left<z\right>= 0.73$ to results at different redshifts from the
literature.  Our last density bin is centred on $log(\Sigma_{10}) =
2.6$ and includes objects up to $log(\Sigma_{10}) = 2.8$, i.e. $\sim
630$ galaxies h$^2$ Mpc$^{-2}$.  With respect to other studies centred
exclusively on rich clusters (e.g. Dressler 1980, Dressler et
al. 1997, Postman et al. 2005), we sample less well the $\sim 1000$
Mpc$^{-2}$ regime, but have a fairly good signal in
intermediate-density regions.  This is indeed expected, looking at the
density map of Figure~\ref{dens_map}.  Overall, our measurement at
$\left<z\right>= 0.73$ is steeper than the local estimate (note that
the abscissa scale is linear): the fraction of early-type galaxies
within this structure at $\left<z\right>=0.73$ goes from about half of
that observed at $z\sim 0$ at low densities ($log(\Sigma_{10}) \sim
1.7$), to $\sim 75\%$ of the local value in the highest density
regime.  In other words, comparing similar environments at the current
epoch and at $z\sim 0.73$ we find a smaller percentage of early-type
galaxies, with this difference being more severe in the lowest density
regime.  Interestingly, Benson et al. (2001, cf. their Figure 6), find
a similar behaviour within their N-body plus semi-analytical
simulations.  In Figure~\ref{morph-dens-comparison}, the agreement of
our result with other high-redshift measurements (Smith et al. 2004,
Postman et al. 2005) is very good for high densities,
$log(\Sigma_{10}) > 2$, while we seem to find a steeper relation at
low densities.  Smith et al. (2004) in particular, who also use
photometric redshifts, seem to have in general a flatter MD relation.

This result is also in good agreement, again for densitites larger than
$log(\Sigma_{10}) \sim 1.5$, with the parallel MD analysis of the
whole COSMOS field using totally independent estimators of galaxy
morphology and local density (cf. Capak et al. 2007b, Figure~6).
Additionally, the approach adopted here is more sensitive to high
densities, nicely extending the Capak et al. measurement for
$0.6<z<0.8$ to densities $100 < \Sigma_{10} < 630$
Mpc$^{-2}$. Conversely, with respect to that measurement our estimated
fraction of early-type galaxies tends to be low for densities below
$log(\Sigma_{10}) \sim 1.5$, either indicating a poor sampling of
low-density environments in our region (as it is indeed the case,
being the sample deliberately centered on a known structure), or being
related to our limits in estimating accurate local densities in these
regimes.

\subsection{Dependence of the morphology-density relation on galaxy
  luminosity}

%
\begin{figure}
\epsscale{1.}
\plotone{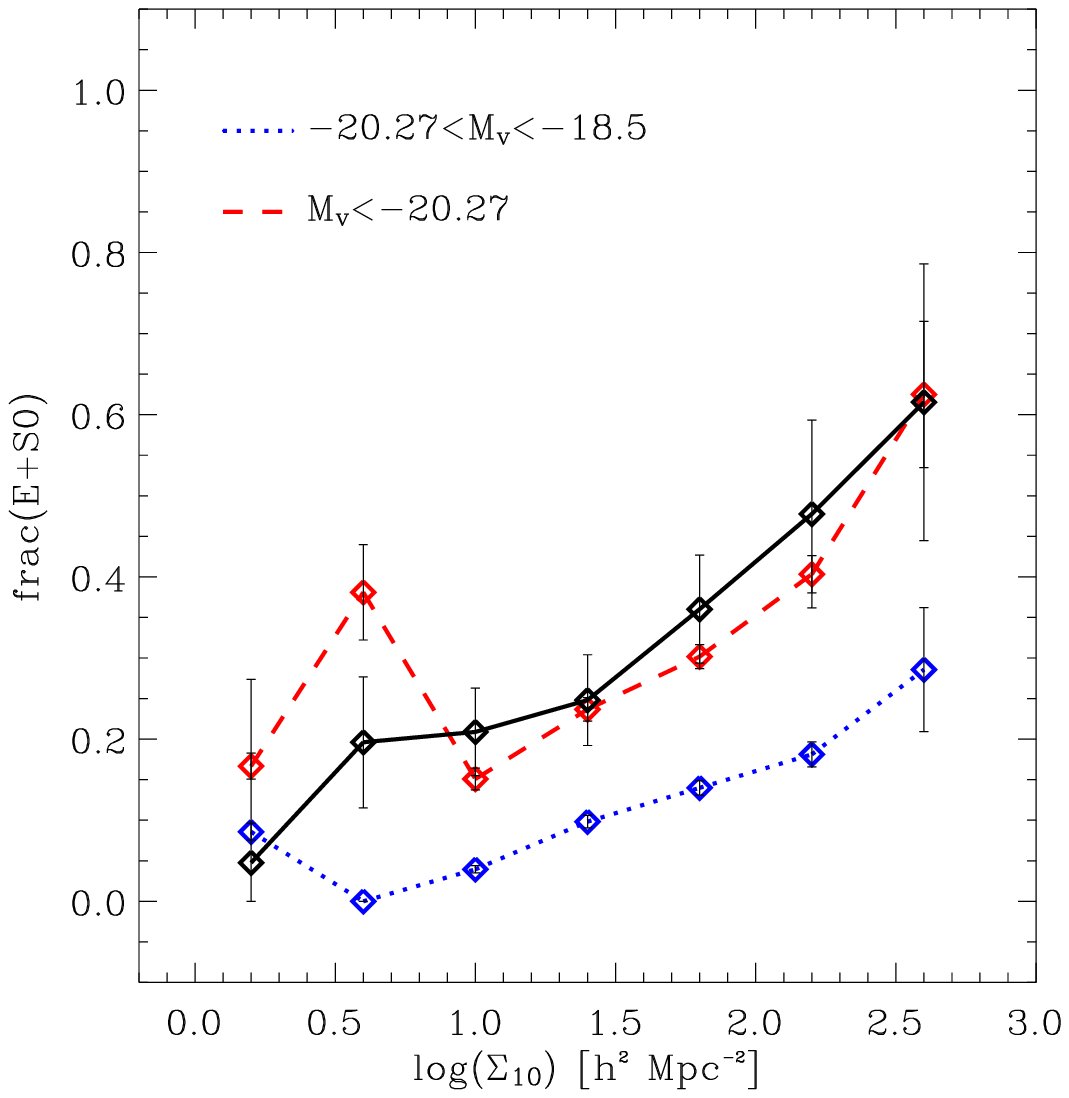}
\caption{The morphology-density relation at $\left<z\right>=0.73$ for
bright and faint galaxies, quantifying the visual impression from
Figure~\ref{morph_map_faint}. The red dashed curve reproduces the
result of Figure~\ref{morph-dens}, i.e. the MD relation for luminous
galaxies with $M_V<-20.27$ (Dressler's limit at this redshift).  The
solid curve gives the same quantity, for the same luminous galaxies,
but now using {\it all} objects in the $I_{AB}<24 $ magnitude-limited
sample as neighbours in the computation of the local density
$\Sigma_{10}$ (this is necessary as to be able to compare bright and
faint galaxies on a consistent basis).  Finally, the dotted blue curve
shows the MD relation for intrinsically faint galaxies with
$-20.27<M_V<-18.5$, again using the whole sample to estimate local density.}
\label{morph-dens-faint}
\end{figure}
\begin{figure}
\epsscale{1.}
\plotone{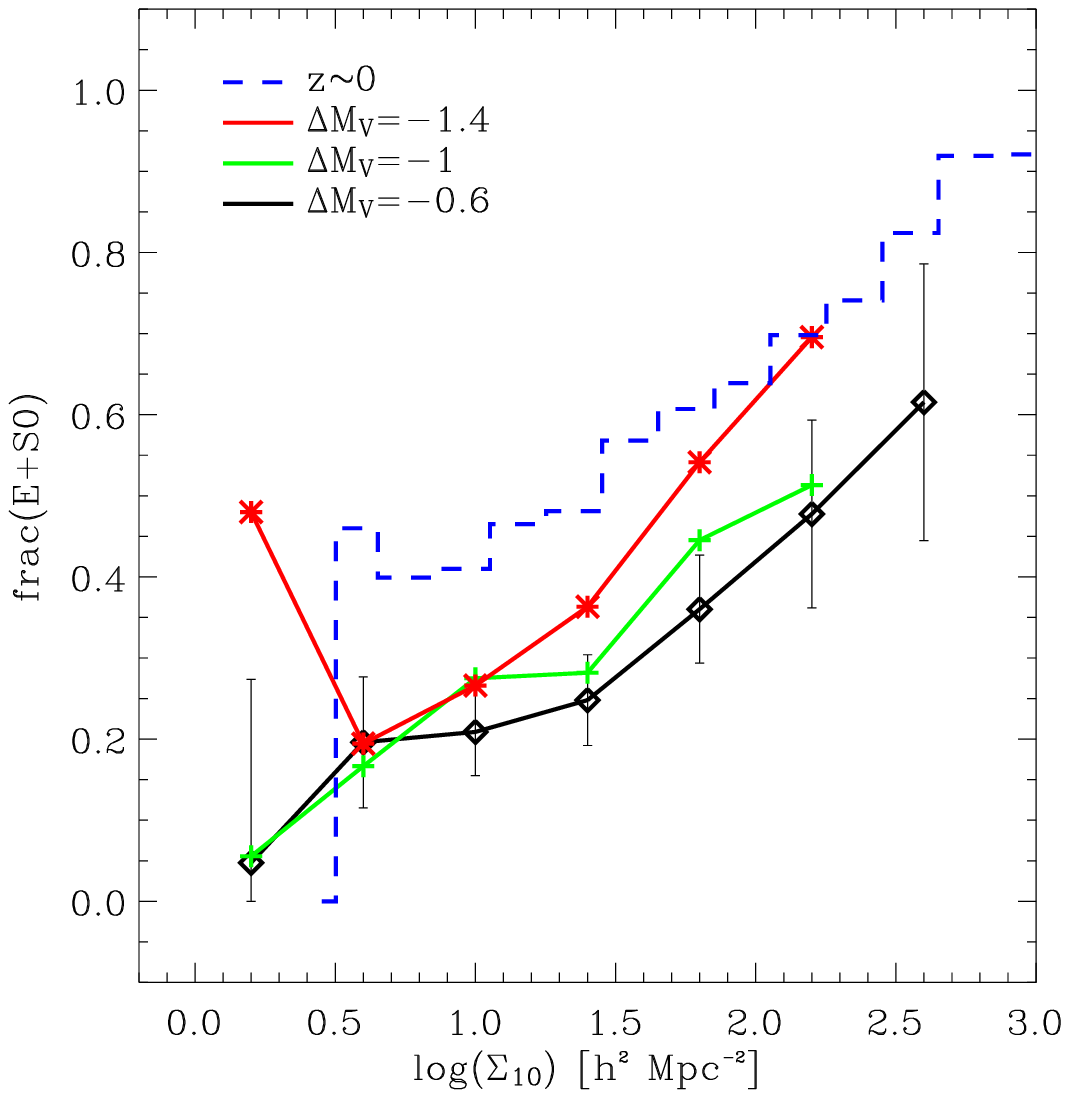}
\caption{Dependence of the measured MD relation on the absolute
magnitude threshold of the galaxy sample analyzed (and thus on the mean
brightening assumed with respect to the $z=0$ measurement) .  $\Delta
m=0.6$ (lower curve) corresponds to our standard choice, adopted to obtain the
results presented throughout the paper, that has been chosen to be
consistent with the latest results on the evolution of the luminosity
function from the VVDS survey (Zucca et al. 2006).  Note how (1) the
measured MD relation is robust even for 
brightening as large as 1 magnitude; (2) only when considering
galaxies brighter than $M_V<-21.7$, the fraction of early-type objects
at densities $\Sigma_{10}=100$ Mpc$^{-2}$ equals that measured at
$z=0$ for $M_V<-19.67$.
}   
\label{brightening}
\end{figure}
%

A quite general prediction of semi-analytical models of galaxy
formation is that the MD relation should show a dependence on
luminosity (e.g. Benson et al. 2001).   Figure~\ref{morph_map_faint}
replicates Figure~\ref{morph_map_bright} but now plotting only
galaxies fainter than Dressler's limit, i.e. $-20.27<M_V<-18.5$, where
the fainter limit corresponds approximately to our apparent magnitude cut
($I_{AB}<24$) at $z=0.73$.
Comparing the two figures
we clearly see that the MD relation is much stronger for luminous
galaxies than for faint ones.
To make this observation quantitative, we 
re-compute the MD relation for this fainter galaxy population.
However, for the comparison between the bright and faint samples to be
meaningful, we need to have a common estimate of local density
$\Sigma_{10}$ for the two samples.  What we need is a measurement
based on the same galaxy background.  We therefore re-estimated the MD
relation for the bright $M_V<-20.27$ sample, but now using the surface
density $\Sigma_{10}$ computed using {\it all} $I_{AB}<24$ galaxies
within the usual $\Delta z = \pm 0.12$ redshift slice.  The result is
shown as a dashed line in Figure~\ref{morph-dens-faint}, with the
solid line repeating the original relationship from
Figure~\ref{morph-dens}.  The comparison of these two curves shows
that a similar MD relation for bright galaxies is consistently detected even
if galaxies of all luminosities are used in the computation of the
background density.  We can therefore compute the MD relation of the
faint sample with respect to this same background, that is is shown by
the bottom dotted line.  Its shallower slope indicates how the MD
relation (at $z\simeq 0.73$) is weakened for galaxies fainter than
Dressler's limit of $M_V = -20.27$.  The simplest interpretation of
this plot is that luminous (massive) early-type galaxies tend to
dominate high-density regions much more than less massive spheroidal
objects.  The latter constitute not more than 30\% of the whole galaxy
population in the highest density bin, about half of the fraction of
luminous early-type galaxies at similar density regimes.

It is also interesting to test the sensitivity of the MD relation to
the level of brightening assumed for Dressler's absolute magnitude threshold.
Figure~\ref{brightening} shows the result of selecting samples which
are respectively 0.6 (our standard choice), 1.0 and 1.4 magnitudes
brighter than Dressler's local value.  As we have discussed in
\S~\ref{proj_dens}, there is some level of ambiguity in choosing
how this luminosity cut-off evolves with redshift and our choice of
$\Delta M_V = -0.6$ is based on the observed mean brightening of $M^*$
from the VVDS (Zucca et al. 2006).  Figure~\ref{brightening} shows
that in fact even for a brightening of 1 magnitude, the measured
relation does not change significantly.   Secondly, the plotted curves
show that only when considering
galaxies brighter than $M_V<-21.7$, the fraction of early-type objects
at densities $\Sigma_{10}=100$ Mpc$^{-2}$ equals that measured at
$z=0$ for $M_V<-19.67$.

\section{Evidence for an influence of cluster ICM phenomena on galaxy morphology} 

%
\begin{figure}
\epsscale{1.}
\plotone{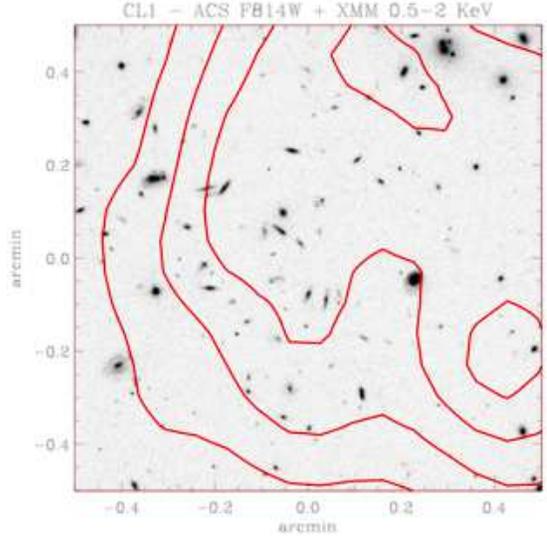}
\caption{ HST-ACS zoom over the X-ray peak, in the area where the
steep gradient in the X-ray surface distribution and the ``hot spot''
in the temperature map are evidence for a shock in the ICM.  Note the
unusual concentration of disk galaxies specifically in this region. }
\label{acs_zoom}
\end{figure}
%

The results obtained in the previous section confirm a strong MD
relation within this large-scale structure at $z\simeq 0.73$, with a
lower fraction of early-type galaxies than at $z=0$, in agreement with
other recent work at high redshift (Smith et al. 2005, Postman et
al. 2005).  In Figure~\ref{acs_zoom} we show the ACS
image of the central region of the main cluster of galaxies, centred
on the X-ray emission peak seen in Figure~\ref{xray_overlay}.  A
sufficiently precise location can be recognized by comparing the X-ray
contours in the two figures (allowing for small changes in their
shape due to differently calculated backgrounds in the two plots).
Immediately at first sight, one notices a surprising number of disk
galaxies dominating the galaxy population in this area.  Most of these
objects have indeed a photometric redshift that places them within the
cluster and their distribution is clearly asymmetric with respect to
the cluster X-ray surface-brightness.  Rather, they tend to
concentrate on the eastern side, in the same region where the steep
gradient in the X-ray profile and the negative jump in the temperature
map (\S~\ref{X-ray}) are found, suggesting a shock in the ICM.  Over
this area, clearly a high-density region, the MD relation seems to
have reversed.  The visual impression is confirmed by our quantitative
morphological classification, that confirms the high fraction of
late-type galaxies in this specific sub-area of the structure, also
when considering objects above Dressler's absolute-magnitude cut
(about two-thirds are brighter than $-20.27$).  This area is not
large enough, however, to significantly impact the MD relation
measured over the entire structure.

\section{Summary and Discussion}

The results obtained in this paper can be summarised as follows:

\begin{itemize}

\item We have identified a large-scale structure at $z\simeq 0.73$ in
the COSMOS survey area, based on a number of independent probes of the
density field.  The surface density distribution of galaxies selected
by photometric redshifts, weak-lensing mass reconstruction and X-ray
surface-brightness distribution show a remarkable agreement in the
overall structure, despite their different selection functions.

\item We have produced different mass estimates of the dominant
  cluster of galaxies, using both weak lensing and X-ray emission.
  Over an aperture of 100 arcsec, the mass from weak lensing is $1.10
  ^{+0.22}_{-0.24} \times10^{14} $ M$_\sun$, while the X-ray mass to
  $r_{500}=84$ arcsec is $M_{500}=1.69_{-0.38}^{+0.58} \times
  10^{14}M_\odot$.
The similarity of these two values within similar apertures is
  encouraging, especially considering the non-full equilibrium
  status of the ICM suggested by the evidence for ongoing merging.  
We have also estimated the total cluster mass by fitting a NFW profile
to the radial shear field, obtaining a best-fitting mass
$M_{NFW}=(6\pm3)\times10^{15}M_\sun$.   

\item The existence of an ongoing major merger between two sub-clusters is
strongly suggested by the temperature map that shows a significant
``ridge'', with
$T_X$ rising to $\sim 10$ keV in the internal region, with respect to
a global measurement of $ \sim 3.51$ keV. This scenario is consistent
with the observed X-ray surface brightness distribution, where a
steeper edge is visible on the eastern side of the cluster. 

\item We have measured the morphology-density relation at
  $\left<z\right>\simeq 0.7$ for a sample of
  more than 1200 galaxies brighter than $M_V=-20.27$.  The resulting
  relation is quite robust with respect to our choice of the
  photometric-redshift slice including the structure. It shows a
  steeper slope with respect to the $z=0$ relation, with about half
  the fraction of early-type galaxies in low-density environments, but
  more than 70\% of the local value within high-density regions.  This
  agrees with general predictions of hierarchical clustering models
  (e.g. Benson et al. 2001), in which the population of early-type
  galaxies in the highest density peaks is established at
  proportionally older epochs.  However, our measurements for
  densities smaller than $\sim 10-30$ Mpc$^{-2}$ have to be treated
  cautiously, given the comparable value of the background correction
  (see Appendix).  For larger values, however, the good agreement with
  recent independent high-redshift estimates is remarkable, indicating
  a reduction in the fraction of early-type galaxies at $z\sim 0.7$ to
  $\sim 75\%$ of the value seen in the local Universe.

\item We have explored the general behaviour of the MD relation at
  magnitudes fainter than the standard value ($-20.27<M_V<-18.5$),
  finding a significantly weaker relationship that corresponds to a
  reduction by a factor of more than two in the fraction of early-type
  galaxies in the highest density bin, between the luminous
  and faint sample.
This trend is generally consistent with predictions from semi-analytic models
(Benson et al. 2001). 
Conversely, our measured ``standard'' MD relation for luminous
galaxies is shown not to be too sensitive to increasing the assumed
intrinsic brightening of galaxies up to $\Delta M_V = 1$ (i.e. for a sample
brighter than $M_V \sim -20.7$).  To actually match the fraction of
early-type galaxies measured at $z\sim 0$ for a density
$\Sigma_{10}=100$ Mpc$^{-2}$, one needs to select only very luminous
objects with $M_V<-21.1$, i.e. $\sim 1.4$ magnitudes brighter.

\item We detect a clear example of how large-scale environmental
  processes such as the merger between groups and sub-clusters indeed
  affects the galaxy population, at least in its apparent morphology.
  In fact, slightly offset with respect to the centre of the main
  cluster, an unusual number of disk galaxies is found, coinciding
  with the colder side of the possible shock front detected in X-rays.
  Several of these galaxies are brighter than Dressler's limit and
  thus do display an ``inverted'' MD relation over this region.  Due
  to the relatively small area of the shock, however, they do not
  affect in a significant way the global morphology-density trend
  measured over the whole structure.
  The natural interpretation of these combined optical and X-ray
  observations is that star formation has been switched-on (or
  enhanced) in these galaxies during the large-scale merger between
  two sub-units, currently mixing together to form the central
  cluster.  To our knowledge, this is the first time that a direct
  link is observed in a high-redshift cluster between the actual
  morphology of galaxies (not only the color or spectroscopic
  properties) and a large-scale shock happening in the ICM surrounding
  them.  It is particularly fascinating to see so many clear, bright
  disks inside a cluster at this redshift.  This implies that disks
  with a significant reservoir of fresh, unprocessed molecular
  hydrogen had to be present in these galaxies before the merging
  event, and have become visible due to the burst of star-formation.
  A connection between the color or star-formation rate of galaxies
  and a merging event has been advocated in the past to explain the
  properties of galaxies in merging clusters.  For example, in the
  core of the Shapley supercluster, where galaxies at the interface of
  two merging clusters are found to have bluer colors (e.g. Bardelli
  et al. 1998).  Similarly, Ferrari et al. (2005) observe an increased
  fraction of emission-line galaxies in the region separating the two
  sub-clusters of A3921, and interpret them as related to the ongoing
  merger.  Even closer to the situation observed here, where we find
  evidence for a merger between two sub-clusters that seems to have
  consequences on the galaxy population, is the case discussed in
  Sakai et al. (2002), Gavazzi et al. (2003) and Cortese et
  al. (2006).  These works discuss the discovery and properties of a
  conspicuous group infalling into the cluster A1367 at 1800
  km~s$^{-1}$, in which twelve members (two giant and ten dwarf
  galaxies) are simultaneously undergoing a conspiquous burst of star
  formation.  This is interpreted as having been produced either by
  tidal interaction among the member galaxies or by ram-pressure by
  the ICM related to the high-speed infall into the cluster.

  In the near future, new multi-wavelength observations of the COSMOS
  field will allow us to explore in more detail the nature of the
  spiral galaxy population in this distant cluster and their
  interaction with the ICM.  In particular, Chandra imaging will
  reveal the details of the shock in the hot gas, while data from IRAC
  and MIPS on board of the Spitzer satellite, will explore directly
  the star formation properties and stellar ages in the spiral
  galaxies.

\end{itemize}

\acknowledgments

We gratefully acknowledge the contributions of the entire COSMOS
collaboration consisting of more than 70 scientists.  The HST COSMOS
Treasury program was supported through NASA grant HST-GO-09822.  The
COSMOS Science Meeting in May 2005 was supported in part by the NSF
through grant OISE-0456439.  This work is based on observations
obtained with XMM--Newton, an ESA science mission with instruments and
contributions directly funded by ESA Member States and the US
(NASA). In Italy, the COSMOS project is supported by INAF under
PRIN-2005/1.06.08.10 and XMM-COSMOS is supported by INAF and MIUR
under grants PRIN/270/2003 and Cofin-03-02-23 and by ASI under grant
ASI/INAF I/023/05/0. In Germany, the XMM--Newton project is supported
by the Bundesministerium f\"ur Bildung und Forschung/Deutsches Zentrum
f\"ur Luft und Raumfahrt, the Max--Planck Society, and the
Heidenhain--Stiftung. Part of this work was supported by the Deutsches
Zentrum f\"ur Luft-- und Raumfahrt, DLR project numbers 50 OR 0207 and
50 OR 0405.
LG thanks Claudio Firmani and Marco Scodeggio for enlightening discussions
on galaxy evolution, Davide Lazzati for invaluable help with IDL
routines and Stefania Giodini for help with the mock samples.

\appendix
\section{Robustness and accuracy of local density estimates}
\label{appendix}

To assess how well we are able to reconstruct
local densities using our photometric redshift catalogue, we performed
a series of tests using one of the mock COSMOS surveys constructed by
Kitzbichler \& White (2006) from the Millennium Simulation.  The
sample we used was tailored as to reproduce the geometry and selection
function of the Subaru data used for the present paper,
i.e. essentially $I_{AB}<24$ for the global sample and $M_V\le -20.27$
for the volume-limited sub-sample with $z=[0.61,0.85]$.  We obtained
very similar results for both the volume-limited and full
magnitude-limited samples.  We show here the results from the latter,
where we have better statistics.

We simulate photometric redshift errors by smoothing the true
redshifts with a Gaussian kernel with $\sigma_z=\sigma_0(1+z)$, where
$\sigma_0=0.03$, i.e. consistent with our estimates from the actual
COSMOS survey (Mobasher et al. 2007).  We are aware that the redshift
error distribution is probably not very well described by a Gaussian,
but based on a number of tests using different $\sigma_0$, we have
seen that this assumption does not affect our conclusions.  At the
position of each galaxy, we shall compare the surface density value
estimated from this simulated photometric-redshift sample
to the corresponding spatial density measured in 3D.  

\begin{figure}
\epsscale{1.}
\plottwo{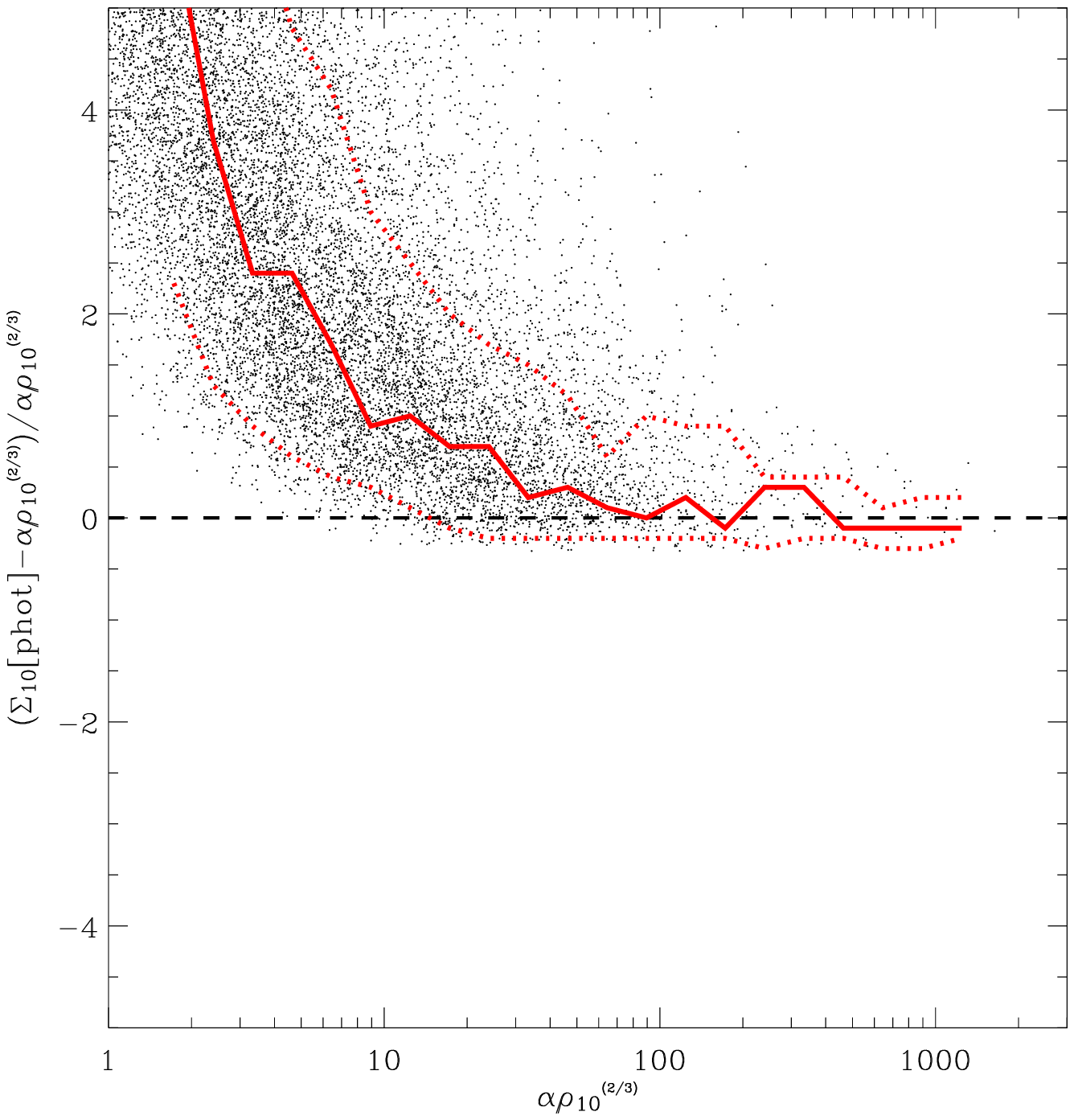}{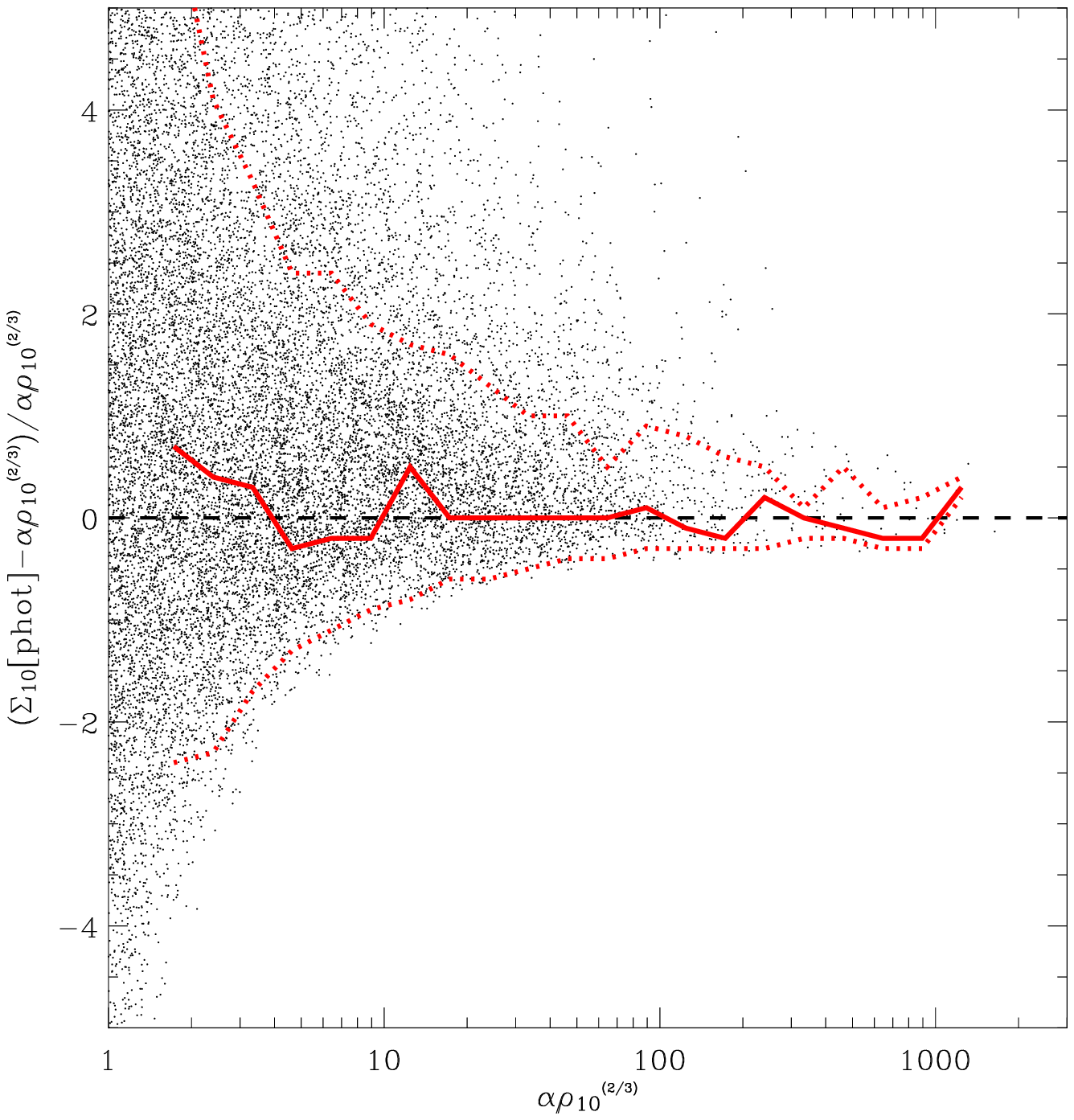}
\caption{Relative difference between projected densities $\Sigma_{10}$
  estimated as discussed in \S~\ref{proj_dens} from a
  ``photometric-redshift'' mock sample and the ``true'' densities one
  would measure with full knowledge of 3D galaxy positions in the same
  sample.  The true density is converted into a projected density as
  indicated.  The left panel shows the comparison without the
  background subtraction described in \S~\ref{proj_dens}, while the
  right panel shows the complete background-subtracted estimates, as
  done on the real data.  The solid line gives the median values
  within bins of constant logarithmic size in true density, while the
  dotted lines enclose the 68.3\% confidence interval around this 
  value.}
\label{app:fig1}
\end{figure}

In Figure~\ref{app:fig1} we plot the normalized differences between
our projected density $\Sigma_{10}$, defined as in \S~\ref{proj_dens},
and estimated from the ``photo-z'' sample and the ``true'' density
measurement one would obtain with perfect knowledge of galaxy
positions in space.  The latter is measured analogously to
$\Sigma_{10}$, but in three-dimensions, i.e. as $\rho_{10}=10\times
3/(4\pi r_{10}^3)$, and then renormalized to obtain the corresponding
2D quantity, $\Sigma_{10}=\alpha\times \rho_{10}^{2/3}$, where
$\alpha=1.76$ is the geometrical factor that accounts for the
projection, assumuing that only galaxies within the original 3D sphere
are involved.  The left panel uses the measurements without
subtracting the mean background, while the right panel shows the
complete background-subtracted estimates, as done on the real data. As
evident, this operation is crucial to reduce the large background
offset introduced at small values of the density by the very thick
slice used to compensate for the large redshift-space blurring. The
solid line gives the expectation values of the measured statistics,
computed as the median value within bins of size $\Delta(\log(\alpha
\rho_{10}^{2/3}))=0.4$, spaced by the same amount along the X axis.
The two dotted lines describe the 68.3\% confidence interval around
this.  

These plots indicate first of all that the subtraction of the
background is important to reduce the bias of our estimator: after the
subtraction, the systematic error of our measurements is consistent
with zero down to local density as small as $\Sigma_{10}=3$ gal h$^2$
Mpc$^{-2}$ (although the probability distribution would still favour
over- rather than under-estimates).  The scatter is clearly very large
at these low densities, however already a galaxy living in an
environment characterized by
10 gal h$^2$ Mpc$^{-2}$, has a 68.3\% probability of being assigned a
density between 0 and 30 gal h$^2$ Mpc$^{-2}$.  Even better, at
densities of 50 gal h$^2$ Mpc$^{-2}$, the 68.3\% error corridor
extends between -0.5 and $\sim +0.8$, so the estimated density will
range between half this value, 25 gal h$^2$ Mpc$^{-2}$, and 1.8 times
it, i.e. 90 gal h$^2$ Mpc$^{-2}$.  The consequence of these errors on
our estimate of the MD relation (Fig.~\ref{morph-dens-comparison}), is
basically that of making the three lowest-density bins at
$\Sigma_{10}\le 10$ h$^{2}$ Mpc$^{-2}$ strongly correlated, shuffling
galaxies among them (consistent
with the nearly flat values found at these regimes). However, above
this value the errors are smaller than the bin size used to estimate
the MD relation.  
These results confirm and corroborate the indications
derived in \S~\ref{proj_dens} from the stability of our estimates as a
function of the slice thickness.  Together with the similar analysis
in the companion paper by Capak et al. (2007), they indicate that current
COSMOS photometric redshifts can be safely used to estimate local
projected densities 
for densities $\Sigma_{10}>5-10$ h$^2$ Mpc$^{-2}$.
Albeit limited in their scope, they also provide in general a
significantly more optimistic view of the ability of photometric-redshift
surveys to study enviromental effects, at least in the regime
of densities where $\Sigma_{10}>10$ h$^{2}$ Mpc$^{-2}$, with respect
to the general conclusions of previous works (e.g. Cooper et
al. 2005). 

%
%
%

\clearpage




\end{document}